# Simulating electron energy-loss spectroscopy and cathodoluminescence for particles in arbitrary host medium using the discrete dipole approximation

Alexander A. Kichigin[1,2] and Maxim A. Yurkin[1,2,*]

[1]*Voevodsky Institute of Chemical Kinetics and Combustion SB RAS, Institutskaya Str. 3, 630090, Novosibirsk, Russia*

[2]*Novosibirsk State University, Pirogova Str. 2, 630090, Novosibirsk, Russia*

*\*Corresponding author: yurkin@gmail.com*

## Abstract

Electron energy-loss spectroscopy (EELS) and cathodoluminescence (CL) are widely used experimental techniques for characterization of nanoparticles. The discrete dipole approximation (DDA) is a numerically exact method for simulating interaction of electromagnetic waves with particles of arbitrary shape and internal structure. In this work we extend the DDA to simulate EELS and CL for particles embedded into arbitrary (even absorbing) unbounded host medium. The latter includes the case of the dense medium, supporting the Cherenkov radiation of the electron, which has never been considered in EELS simulations before. We build a rigorous theoretical framework based on the volume-integral equation, final expressions from which are implemented in the open-source software package ADDA. This implementation agrees with both the Lorenz-Mie theory and the boundary-element method for spheres in vacuum and moderately dense host medium. And it successfully reproduces the published experiments for particles encapsulated in finite substrates. The latter is shown for both moderately dense and Cherenkov cases – a gold nanorod in $SiO_2$ and a silver sphere in $SiN_x$, respectively.

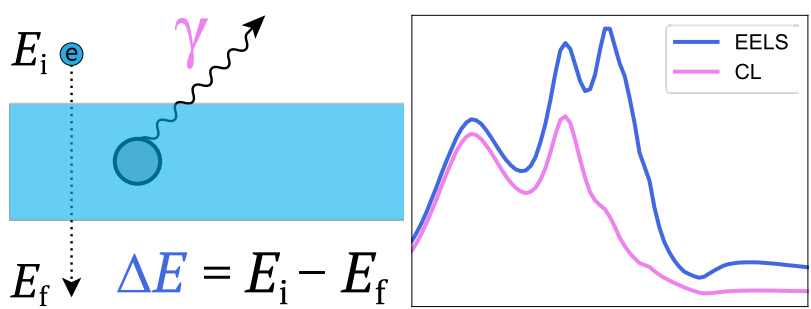

# 1  Introduction

Optical excitations of small objects have been gaining interest over the past decades. The quest for the highest possible space-energy resolution is unachievable by purely optical methods due to the diffraction limit.[1] This limit can be overcome using fast electrons as a probe instead of light – the corresponding experimental technique is called the electron energy-loss spectroscopy (EELS). EELS is a well-developed extension of the standard electron microscope,[2] where a particle under study is exposed to a beam of relativistic electrons with the same kinetic energies of the order of 100 keV. The energies of transmitted electrons are measured leading to the energy-loss spectrum (EELS spectrum). Currently, EELS provides information about excitations with sub-Å and few-meV space-energy resolution.[3] While exposed to an electron beam, the particle emits photons – this phenomenon, called cathodoluminescence (CL), also provides information about photonic properties.[4] To accurately interpret the results of an EELS or CL experiment, it is necessary to have a theoretical description of particle interaction with the electromagnetic field of a fast electron, complemented by a simulation method.

There are many methods capable of simulating EELS, from the analytical Lorenz-Mie theory for spheres[5] to surface-discretization boundary-element method (BEM)[6] and volume-discretization methods, such as finite-difference time-domain[7] and the discrete-dipole approximation (DDA).[8,9] While a theoretical description of EELS can be developed in a general setting of a particle placed in arbitrary infinite host medium, most of the numerical methods apply only to vacuum environment. The notable exception is MNPBEM,[10] which seems to support arbitrary infinite host medium by its internal parameters. However, this option is poorly documented and the corresponding simulations have been performed only for the case of non-absorbing medium with relatively low density.[11,12]

Thus, the simulation capabilities do not fully support the experimental conditions, where the particle is always placed on or inside a substrate, or require discretizing a large chunk of substrate in addition to the particle itself. The presence of a substrate affects the EELS/CL spectra.[13] e.g., by redshifting localized surface plasmon resonances (LSPRs) with increasing host-medium refractive index.[14] Although the development of very thin substrates has made the experimental results close to vacuum simulations,[15,13] some experimental studies intentionally consider particles inside a substrate.[16,12,17] Importantly, sufficiently dense host medium may slow down the light below the electron speed, leading to qualitatively different case of Cherenkov radiation.[16] This case may become more common with increasing electron energy, but has never been considered in numerical simulations of EELS.

The goal of this paper is to enhance the capabilities of EELS/CL simulations using the DDA – a numerically exact method for simulating interaction of electromagnetic waves with particles of arbitrary shape and internal structure,[18] based on the volume integral equation (VIE) in the frequency domain.[19] The popularity of the DDA is based on conceptual simplicity, combined with availability of two open-source highly-optimized codes: DDSCAT[20] and ADDA.[21] Initially designed for simulating interaction of particles with plane waves,[20] the DDA is applicable to arbitrary electromagnetic fields,[18] including the field of a fast electron. The latter has been exemplified by two specialized codes: DDEELS[8] and e-DDA,[9] which, however, assume the vacuum surrounding. Moreover, the underlying VIE-based theory of EELS is available only for this case.[8,9,22]

In this paper we, first, construct a general theoretical framework of EELS and CL for particles in arbitrary host medium (Section 2), based on the VIE and energy-budget considerations.[23] When the host medium is absorbing or incurs Cherenkov radiation, particle-induced energy losses are accompanied by free-space ones. We discuss the feasibility of separating these two losses in experimental signals in Sections 2.5 and 2.6. Moreover, we extend the scale invariance rule of electromagnetics to EELS simulations, expanding the applicability domain of existing vacuum-based codes in Section 2.7.

Second, we implement the obtained general expressions in the ADDA code together with a dedicated Python library, allowing wide range of EELS and CL simulations in arbitrary host medium out of the box (Section 3.1). Next, we demonstrate the high accuracy of this implementation in comparison with the Lorenz-Mie theory and MNPBEM for a test sphere in vacuum (Section 3.2) and various host media (Section 3.3). The latter comparison highlights the limitations of MNPBEM in Cherenkov case. Finally, Section 3.4 demonstrates the practical applicability of the updated ADDA code by accurately reproducing experimental EELS data for a silver nanosphere encapsulated in non-absorbing Cherenkov host medium[16] and for a nanowire placed inside glass substrate.[12] We use SI units throughout the manuscript, and preliminary results of this paper have been presented in [24,25].

## 2 Theory

### 2.1 Energy budget for time-harmonic sources

This section is based on [19,23,26]. Let us define the current density of time-harmonic external sources $\mathbf{J}_s(\mathbf{r})$ to be independent of the resulting electromagnetic field and consider nonmagnetic ($\mu = \mu_0$) isotropic host medium with dielectric permittivity $\varepsilon_h = m_h^2 \varepsilon_0$ ($m_h$ is its refractive index). The incident (or "source-generated") electromagnetic field must satisfy the Maxwell equations in $\mathbb{R}^3$:

$$\begin{aligned}\nabla \times \mathbf{E}_{\text{inc}}(\mathbf{r}) &= i\omega\mu_0 \mathbf{H}_{\text{inc}}(\mathbf{r}), \\ \nabla \times \mathbf{H}_{\text{inc}}(\mathbf{r}) &= -i\omega\varepsilon_h \mathbf{E}_{\text{inc}}(\mathbf{r}) + \mathbf{J}_s(\mathbf{r}).\end{aligned} \quad (1)$$

Below we consider the cases of non-absorbing ($\varepsilon_h \in \mathbb{R}$) and absorbing ($\varepsilon_h \in \mathbb{C}$) host media. In both cases we assume $0 \leq \arg \varepsilon_h < \pi$ (passive medium) and therefore $0 \leq \arg m_h < \pi/2$, where arg is the complex argument, for which we assume the range $(-\pi, \pi]$.

A particle is a nonmagnetic object with finite volume $V_{\text{int}}$ and complex isotropic permittivity distribution $\varepsilon_p(\mathbf{r}) = m_p^2(\mathbf{r})\varepsilon_0$ ($m_p(\mathbf{r})$ is the particle's refractive index). Then the dielectric permittivity function in the whole space is

$$\varepsilon(\mathbf{r}) \stackrel{\text{def}}{=} \begin{cases} \varepsilon_h, & \mathbf{r} \in V_{\text{ext}}, \\ m^2(\mathbf{r})\varepsilon_h, & \mathbf{r} \in V_{\text{int}}, \end{cases} \quad (2)$$

where $V_{\text{ext}} = \mathbb{R}^3 \setminus V_{\text{int}}$, $m(\mathbf{r})$ is the refractive index *relative to the host medium*: $m(\mathbf{r}) \stackrel{\text{def}}{=} m_p(\mathbf{r})/m_h$.

The presence of the particle changes the electromagnetic field in the whole space, composed of the particle's volume $V_{\text{int}}$ and the external volume $V_{\text{ext}}$. This field must satisfy the following Maxwell's equations:

$$\begin{aligned}\nabla \times \mathbf{E}(\mathbf{r}) &= i\omega\mu_0 \mathbf{H}(\mathbf{r}), \\ \nabla \times \mathbf{H}(\mathbf{r}) &= -i\omega\varepsilon(\mathbf{r})\mathbf{E}(\mathbf{r}) + \mathbf{J}_s(\mathbf{r}),\end{aligned} \quad (3)$$

and boundary conditions at the particle interface. These equations are equivalent to the following volume integral equation (VIE) for the electric field $\mathbf{E}(\mathbf{r})$:[19]

$$\mathbf{E}(\mathbf{r}) = \mathbf{E}_{\text{inc}}(\mathbf{r}) + k^2 \lim_{V_0 \to 0} \int_{\mathbb{R}^3 \setminus V_0} d^3 r' [m^2(\mathbf{r}') - 1]\overline{\mathbf{G}}(\mathbf{r}, \mathbf{r}') \cdot \mathbf{E}(\mathbf{r}') - \frac{m^2(\mathbf{r}) - 1}{3}\mathbf{E}(\mathbf{r}), \quad (4)$$

where $V_0$ is a sphere centered at $\mathbf{r}$ to exclude the singularity, $k = \omega\sqrt{\varepsilon_h \mu_0} = m_h k_0$ is the wavenumber in the host medium ($c$ is the speed of light in vacuum, and $k_0 = \omega/c$ is the vacuum wavenumber), $\overline{\mathbf{G}}(\mathbf{r}, \mathbf{r}')$ is the free-space Green's tensor, defined as

$$\overline{\mathbf{G}}(\mathbf{r}, \mathbf{r}') \stackrel{\text{def}}{=} \left(\overline{\mathbf{I}} + \frac{\nabla \otimes \nabla}{k^2}\right) \frac{\exp(ik|\mathbf{r} - \mathbf{r}'|)}{4\pi|\mathbf{r} - \mathbf{r}'|}, \quad (5)$$

where $\overline{\mathbf{I}}$ is the identity tensor, and $\otimes$ denotes dyadic (tensor) product.

Incident electric field can be expressed in terms of $\mathbf{J}_s(\mathbf{r})$ as

$$\mathbf{E}_{\text{inc}}(\mathbf{r}) = i\omega\mu_0 \lim_{V_0 \to 0} \int_{V_s \setminus V_0} d^3 r' \overline{\mathbf{G}}(\mathbf{r}, \mathbf{r}') \cdot \mathbf{J}_s(\mathbf{r}') - i\frac{\mathbf{J}_s(\mathbf{r})}{3\omega\varepsilon_h}, \quad (6)$$

where $V_s$ is the volume enclosing the sources, and the exclusion of the singularity in the volume $V_0$ makes the expression valid in the whole $\mathbb{R}^3$. In this work we assume that the sources are outside of the particle: $V_s \cap V_{\text{int}} = \emptyset$ (unless noted otherwise). Let us further define the polarization density of the particle as

$$\mathbf{P}(\mathbf{r}) = [\varepsilon(\mathbf{r}) - \varepsilon_h]\mathbf{E}(\mathbf{r}), \tag{7}$$

then the scattered electric field is expressed as

$$\mathbf{E}_{\text{sca}}(\mathbf{r}) \stackrel{\text{def}}{=} \mathbf{E}(\mathbf{r}) - \mathbf{E}_{\text{inc}}(\mathbf{r}) = \omega^2 \mu_0 \lim_{V_0 \to 0} \int_{V_{\text{int}} \setminus V_0} d^3 r' \overline{\mathbf{G}}(\mathbf{r}, \mathbf{r}') \cdot \mathbf{P}(\mathbf{r}') - \frac{\mathbf{P}(\mathbf{r})}{3\varepsilon_h}. \tag{8}$$

Time-averaged electromagnetic-energy transfer per unit area is given by the Poynting vector [27]:

$$\mathbf{S}(\mathbf{r}) = \frac{1}{2} \operatorname{Re}[\mathbf{E}(\mathbf{r}) \times \mathbf{H}^*(\mathbf{r})]. \tag{9}$$

Integrating it over the closed surface $A$ results in the power generated or lost in the volume inside the surface (according to the Poynting theorem):

$$W = \oint_A d\mathbf{A} \cdot \mathbf{S}(\mathbf{r}), \tag{10}$$

where $d\mathbf{A} \stackrel{\text{def}}{=} \mathbf{n}\, d^2 r$, $\mathbf{n}$ is the vector normal to the surface, and the sign is chosen such that $W$ is positive when energy goes outside the surface. Using the divergence theorem, the surface integral is transformed into the volume integral:

$$W = -\frac{\omega}{2} \int_{V_A} d^3 r |\mathbf{E}(\mathbf{r})|^2 \operatorname{Im}[\varepsilon(\mathbf{r})] - \frac{1}{2} \int_{V_A} d^3 r \operatorname{Re}[\mathbf{E}(\mathbf{r}) \cdot \mathbf{J}_s^*(\mathbf{r})]. \tag{11}$$

This expression is valid only if $\mathbf{E}(\mathbf{r})$ and $\mathbf{J}_s(\mathbf{r})$ are square-integrable inside $V_A$. Note that the dot product of two vectors does *not* imply conjugation of the second operand. This is consistent with the definition of action of tensor on the vector, as used above. In other words, this dot product is not a proper inner product of two complex vectors.

For the case of non-absorbing host medium ($m_h \in \mathbb{R}$), let us apply Eq. (11) to $V_s$ in the absence of the particle. Substituting $\mathbf{E}_{\text{inc}}(\mathbf{r})$ for $\mathbf{E}(\mathbf{r})$ gives us the free-space energy loss power $W_0$. The first component of Eq. (11) equals zero, and the second one can be rewritten as:[23]

$$W_0 = \frac{\omega \mu_0}{2} \iint_{V_s} d^3 r\, d^3 r'\, \mathbf{J}_s^*(\mathbf{r}) \cdot \overline{\mathbf{G}}^I(\mathbf{r}, \mathbf{r}') \cdot \mathbf{J}_s(\mathbf{r}'), \tag{12}$$

where we use the notation

$$\overline{\mathbf{G}}^I(\mathbf{r}, \mathbf{r}') \stackrel{\text{def}}{=} \frac{1}{2i} \{\overline{\mathbf{G}}(\mathbf{r}, \mathbf{r}') - [\overline{\mathbf{G}}(\mathbf{r}', \mathbf{r})]^H\}, \tag{13}$$

which is a symmetric (self-adjoint) operator kernel, and $H$ denotes the Hermitian (conjugate) transpose of a tensor (matrix). In isotropic medium

$$\overline{\mathbf{G}}^I(\mathbf{r},\mathbf{r}') = \mathrm{Im}[\overline{\mathbf{G}}(\mathbf{r},\mathbf{r}')], \qquad (14)$$

and in non-absorbing medium $\overline{\mathbf{G}}^I(\mathbf{r},\mathbf{r}')$ is always finite,[23] in particular:

$$\lim_{\mathbf{r}\to\mathbf{r}'} \overline{\mathbf{G}}^I(\mathbf{r},\mathbf{r}') = \frac{k\overline{\mathbf{I}}}{6\pi}. \qquad (15)$$

Then, as explained in [23], Eq. (12) is continuous versus $\mathbf{J}_s(\mathbf{r})$, whenever the latter is integrable. Thus, it is also valid for delta-functions although they are not square-integrable.

By applying Eq. (11) to $V_s$ and $\mathbf{E}_{\mathrm{sca}}$ in a similar manner, we obtain the particle-induced energy loss power $W_{\mathrm{enh}}$, while the integral over $V_{\mathrm{int}}$ gives the extinction power (a standard quantity in light scattering problems):[23]

$$\begin{aligned} W_{\mathrm{enh}} &\stackrel{\mathrm{def}}{=} -\frac{1}{2} \int_{V_s} d^3\mathbf{r}\, \mathrm{Re}[\mathbf{E}_{\mathrm{sca}}(\mathbf{r}) \cdot \mathbf{J}_s^*(\mathbf{r})] \\ &= -\frac{\omega^2 \mu_0}{2} \int_{V_s} d^3\mathbf{r} \int_{V_{\mathrm{int}}} d^3\mathbf{r}'\, \mathrm{Re}[\mathbf{J}_s^*(\mathbf{r}) \cdot \overline{\mathbf{G}}(\mathbf{r},\mathbf{r}') \cdot \mathbf{P}(\mathbf{r}')], \end{aligned} \qquad (16)$$

$$\begin{aligned} W_{\mathrm{ext}} &\stackrel{\mathrm{def}}{=} -\frac{\omega}{2} \int_{V_{\mathrm{int}}} d^3\mathbf{r}\, \mathrm{Im}[\mathbf{E}_{\mathrm{inc}}(\mathbf{r}) \cdot \mathbf{P}^*(\mathbf{r})] \\ &= -\frac{\omega^2 \mu_0}{2} \int_{V_{\mathrm{int}}} d^3\mathbf{r} \int_{V_s} d^3\mathbf{r}'\, \mathrm{Re}[\mathbf{P}^*(\mathbf{r}) \cdot \overline{\mathbf{G}}(\mathbf{r},\mathbf{r}') \cdot \mathbf{J}_s(\mathbf{r}')]. \end{aligned} \qquad (17)$$

For completeness, we also provide the expression for absorbed power:[23]

$$W_{\mathrm{abs}} \stackrel{\mathrm{def}}{=} \frac{\omega}{2} \int_{V_{\mathrm{int}}} d^3\mathbf{r}\, \mathrm{Im}[\varepsilon(\mathbf{r})] |\mathbf{E}(\mathbf{r})|^2. \qquad (18)$$

Schematically these and other powers are visualized in Fig. 1, with the only caveat that we further consider the current source corresponding to a moving electron, for which $V_s$ is an infinite line.

Interestingly, Eqs. (16)–(18) can also be used in the case of absorbing host medium, but their physical meaning is ambiguous. Absorption by the host medium significantly changes the whole energy budget, particularly positions of the contours in Fig. 1 cannot be freely moved around the particle and the sources. This problem does not have a fully satisfactory solution even for the interpretation of $W_{\mathrm{ext}}$ in the standard scattering problem for the plane electromagnetic waves.[28,29] A similar ambiguity appears if the source is placed inside the particle, i.e. $V_s$ and $V_{\mathrm{int}}$ overlap. We leave a rigorous analysis of this case for future research, but it must include certain assumption that sources and the medium are physically separated, i.e. $V_s$ is effectively excluded from $V_{\mathrm{int}}$. In the case of delta-function source (point, line, or surface) this will additionally incur a small exclusion volume, similar to that used in Eq. (8), but with the shape corresponding to that of the source (spherical, cylindrical, or planar).

Moreover, if the medium around the sources is absorbing, some of the results may depend on the width of this exclusion volume, not having a bounded zero limit. The corresponding problem of defining $W_0$ in absorbing host medium is discussed in Section 2.3.

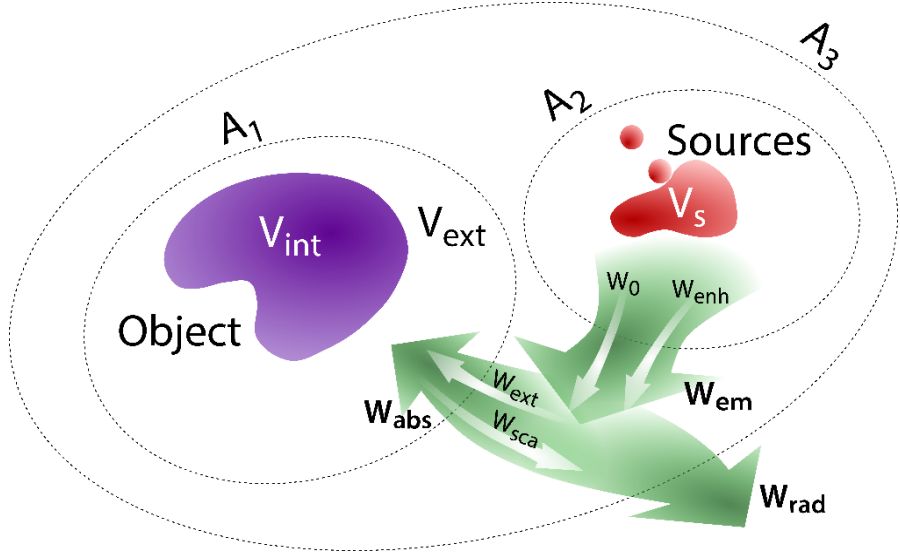

Fig. 1. Visualization of energy powers in the energy budget framework. Reproduced from [23].

It is worth mentioning that cross-sections $C$ (extinction, absorption, scattering) are commonly used in scattering problems for the plane waves. They are expressed in terms of the corresponding power $W$ and the intensity of the incident wave $I_0$:[30]

$$C_X = \frac{W_X}{I_0}, \qquad I_0 \stackrel{\text{def}}{=} m'_\text{h} \frac{\varepsilon_0 c}{2} E_0^2, \tag{19}$$

where $X$ is a subscript of any power from Fig. 1, $m'_\text{h}$ is the real part of $m_\text{h}$, and $E_0$ is the amplitude of the incident wave. $E_0$ has physical meaning for the plane waves, and can be meaningfully defined for some other shaped beams. Moreover, postulating any constant instead of $E_0$, e.g., unity multiplied by appropriate unit, allows one to define all cross-sections for the moving electron as well. This can be convenient, since they are expressed in units of area and, thus, can be trivially converted between different systems of units.

## 2.2 Electric field of a relativistic electron

We consider a relativistic electron as a point charge $q$ moving with the speed $v$ in the positive direction of the $z$-axis. At the time $t = 0$ the charge has coordinates $\mathbf{r}_0 = (x_0, y_0, z_0)$. The corresponding current density is

$$\mathbf{J}_\text{s}(\mathbf{r}, t) = qv\delta(x - x_0)\delta(y - y_0)\delta(z - z_0 - vt)\mathbf{e}_z, \tag{20}$$

where $\mathbf{e}_z$ is a unit vector along the $z$-axis. After applying the Fourier transform (defined as Eq. (S1) in the Supporting Information) we obtain the current density in the frequency domain

$$\mathbf{J}_s(\mathbf{r}) = q\delta(x-x_0)\delta(y-y_0)\exp\left[\mathrm{i}\frac{\omega}{v}(z-z_0)\right]\mathbf{e}_z. \tag{21}$$

To find the incident electric field, we substitute Eq. (21) into Eq. (6) for arbitrary host medium ($m_\mathrm{h} \in \mathbb{C}$), dependence on which through $k$ is implicit in the Green's tensor:

$$\mathbf{E}_{\mathrm{inc}}(\mathbf{r}) = \mathrm{i}\omega\mu_0\left(\bar{\mathbf{I}} + \frac{\nabla\otimes\nabla}{k^2}\right)\cdot\int_{V_s} \mathrm{d}^3\mathbf{r}' \frac{\exp(\mathrm{i}k|\mathbf{r}-\mathbf{r}'|)}{4\pi|\mathbf{r}-\mathbf{r}'|}\mathbf{J}_s(\mathbf{r}') = \mathrm{i}\omega\mu_0\left(\mathbf{e}_z + \frac{\nabla}{k^2}\frac{\partial}{\partial z}\right)I_1(\mathbf{r}), \tag{22}$$

where changing the order of integration and differentiation eliminates the second term in Eq. (6) and makes the singularity of the kernel integrable (the exclusion volume is then redundant).[31] The remaining integral is evaluated in Section S3 of the Supporting Information:

$$I_1(\mathbf{r}) \stackrel{\text{def}}{=} \int_{V_s} \mathrm{d}^3\mathbf{r}' \frac{\exp(\mathrm{i}k|\mathbf{r}-\mathbf{r}'|)}{4\pi|\mathbf{r}-\mathbf{r}'|}J_{s,z}(\mathbf{r}') = \frac{q}{2\pi}\exp\left[\mathrm{i}\frac{\omega}{v}(z-z_0)\right]K_0\left(\frac{\omega b}{\gamma_\mathrm{h} v}\right), \tag{23}$$

where we introduced $b \stackrel{\text{def}}{=} \sqrt{(x-x_0)^2 + (y-y_0)^2}$ and the notation analogous to the one used in the special relativity theory:

$$\beta_\mathrm{h} \stackrel{\text{def}}{=} \frac{v}{c}m_\mathrm{h}, \qquad \gamma_\mathrm{h} \stackrel{\text{def}}{=} \sqrt{\frac{1}{1-\beta_\mathrm{h}^2}}. \tag{24}$$

The principal branch for the square-root function is chosen such that it continuously depends on $\varepsilon_\mathrm{h}$ when $0 \leq \arg\varepsilon_\mathrm{h} < \pi$ (or, equivalently, $0 \leq \arg m_\mathrm{h} < \pi/2$ which we assumed in Section 2.1), except for the singularity at $\beta_\mathrm{h} = 1$. Then $\gamma_\mathrm{h}$ lies in the first quadrant of the complex plane ($0 \leq \arg\gamma_\mathrm{h} \leq \pi/2$) excluding the interval $[0,1)$. Specifically, $\beta_\mathrm{h} < 1$ and $\beta_\mathrm{h} > 1$ lead to real ($\gamma_\mathrm{h} > 1$) and imaginary ($\mathrm{Im}\,\gamma_\mathrm{h} > 0$) $\gamma_\mathrm{h}$, respectively, corresponding to relatively less and more dense non-absorbing host media. In the case of vacuum, we always have $\beta_\mathrm{h} < 1$.

We substitute Eq. (23) into Eq. (22) and obtain:

$$\mathbf{E}_{\mathrm{inc}}(\mathbf{r}) = \frac{q\omega}{2\pi\varepsilon_0 m_\mathrm{h}^2 v^2 \gamma_\mathrm{h}}\exp\left[\mathrm{i}\frac{\omega}{v}(z-z_0)\right]\begin{pmatrix} \frac{x-x_0}{b}K_1\left(\frac{\omega b}{\gamma_\mathrm{h} v}\right) \\ \frac{y-y_0}{b}K_1\left(\frac{\omega b}{\gamma_\mathrm{h} v}\right) \\ -\frac{\mathrm{i}}{\gamma_\mathrm{h}}K_0\left(\frac{\omega b}{\gamma_\mathrm{h} v}\right) \end{pmatrix}. \tag{25}$$

where we used $K_0'(z) = -K_1(z)$ (Eq. 10.29.3 of [32]). The behavior of the incident field is mostly determined by $\gamma_\mathrm{h}$. When it has a positive real part ($\mathrm{Re}\,\gamma_\mathrm{h} > 0$), the field decays exponentially with $b$ (Eq. 10.25.3 of [32]). By contrast, when $\gamma_\mathrm{h}$ is purely imaginary (corresponding to $\beta_\mathrm{h} > 1$), the field oscillates and decays as $1/\sqrt{b}$ similar to the field of a line source.

To conclude, we derived the field in the most general case of an arbitrary host medium ($m_\text{h} \in \mathbb{C}$). This expression includes the case of non-absorbing host medium ($m_\text{h} \in \mathbb{R}$) or vacuum as special cases, and allow straightforward quasistatic limit. In all cases the final results match the ones described in the literature.[33] However, the latter were derived using different approaches for solving the Maxwell equations for different cases of the host media. By contrast, we used a single VIE framework with only technical differences in evaluating integrals. The observed agreement, thus, supports the universal applicability of this framework, which we further use to evaluate interaction of the electron field with a nanoparticle.

## 2.3 Free-space energy losses

To find the free-space energy losses of a relativistic electron moving in an infinite non-absorbing host medium ($m_\text{h} \in \mathbb{R}$), we substitute Eq. (21) into Eq. (12):

$$W_0 = \frac{\omega \mu_0 q^2}{8\pi} \iint_{-\infty}^{\infty} \text{d}z \text{d}z' \exp\left[i\frac{\omega}{v}(z'-z)\right]\left(1 + \frac{1}{k^2}\frac{\partial^2}{\partial z^2}\right)\frac{\sin[k(z'-z)]}{z'-z}$$

$$\Rightarrow \frac{\partial W_0}{\partial z} = \frac{\omega \mu_0 q^2}{8\pi} I_2(\beta_\text{h}),$$

(26)

where we used $k \in \mathbb{R}$ in evaluating $\overline{\mathbf{G}}^I(\mathbf{r}, \mathbf{r}')$, expressed the loss power per unit distance, and denoted the remaining one-dimensional integral as $I_2(\beta_\text{h})$. To calculate it we replace $z$ with $z'$ in the derivative and substitute $u \stackrel{\text{def}}{=} \omega(z'-z)/v$:

$$\begin{aligned}
I_2(\beta_\text{h}) &\stackrel{\text{def}}{=} \int_{-\infty}^{\infty} \text{d}u \exp(iu)\left(1 + \frac{1}{\beta_\text{h}^2}\frac{\partial^2}{\partial u^2}\right)\frac{\sin(\beta_\text{h} u)}{u} \\
&= \left(1 - \frac{1}{\beta_\text{h}^2}\right)\int_{-\infty}^{\infty} \text{d}u \exp(iu) \frac{\sin(\beta_\text{h} u)}{u} \\
&= \frac{1}{\beta_\text{h}^2 \gamma_\text{h}^2}\int_0^{\infty} \frac{\text{d}u}{u}\{\sin[(\beta_\text{h}-1)u] + \sin[(\beta_\text{h}+1)u]\} \\
&= \frac{\pi}{2\beta_\text{h}^2 \gamma_\text{h}^2}[\text{sgn}(\beta_\text{h}-1) + 1],
\end{aligned}$$

(27)

where we integrated by parts twice, and used the known limit for the sine integral. Note that the integrals converge at infinity due to the oscillating kernel (since $\beta_\text{h} \neq 1$) decreasing as $1/u$, and the integrand is regular at zero (together with its derivatives).

By substituting the result for $I_2$ into Eq. (26), we obtain that in vacuum or in a medium with $\beta_\text{h} < 1$ the free-space energy loss power is

$$W_0 = 0,$$

(28)

which is a known fact: a charge moving slower than the speed of light in the medium does not lose energy. For $\beta_\text{h} > 1$, when the speed of charge exceeds the speed of light in the medium,

$$\frac{\partial W_0}{\partial z} = \frac{\mu_0}{8} q^2 \omega \left(1 - \frac{c^2}{v^2 m_h^2}\right), \tag{29}$$

which vanish when $\beta_h \to 1$. Hence, the validity of Eqs. (28) and (29) can be extended by continuity to $\beta_h \leq 1$ and $\beta_h \geq 1$, respectively. Finally, the free-space energy loss per unit distance, given by Eq. (S6), is

$$\frac{\partial}{\partial z}\Delta E_0 = \frac{\mu_0}{4\pi} q^2 \int_{\beta_h(\omega)>1} d\omega\, \omega \left(1 - \frac{c^2}{v^2 m_h^2}\right). \tag{30}$$

Thus, we obtained the well-known Frank-Tamm formula[34] for Cherenkov radiation[35] in any non-absorbing host medium, which again shows the versatility of our theoretical framework.

In an infinite absorbing host medium ($m_h \in \mathbb{C}$) the integral in the expression for $W_0$ [Eq. (12)] will become singular due to the divergence of $\overline{\mathbf{G}}^I(\mathbf{r}, \mathbf{r}')$ for $\mathbf{r} \to \mathbf{r}'$. As discussed in Section 2.1 the possible workaround is to assume a finite exclusion volume in the host medium around the electron trajectory, which is equivalent to assuming a finite electron size.[33] In principle, the latter provides practically usable expressions, but we do not discuss it in this paper.

Note, however, that, conceptually, this singularity indicates the limitation of macroscopic Maxwell's equations. Since the latter are obtained by averaging microscopic ones over the size of several atomic scales, they are not expected to be accurate at smaller scales. And it is exactly the latter scales that become important for electron losses in absorbing medium. Microscopically, the electron is always moving in the vacuum, thus the losses (per unit of length) are always finite.

## 2.4 Particle-induced energy losses

To express particle-induced energy losses, we start from Eq. (16) considering the most general host medium ($m_h \in \mathbb{C}$). We use the general property of tensor transposition

$$\forall \mathbf{a}, \mathbf{b}: \mathbf{a} \cdot \overline{\mathbf{A}} \cdot \mathbf{b} = \mathbf{b} \cdot \overline{\mathbf{A}}^T \cdot \mathbf{a}, \tag{31}$$

and Green's tensor's symmetry (reciprocity) $\overline{\mathbf{G}}(\mathbf{r}, \mathbf{r}') = [\overline{\mathbf{G}}(\mathbf{r}', \mathbf{r})]^T$ to change the order of integration:

$$W_{\text{enh}} = -\frac{\omega}{2} \int_{V_{\text{int}}} d^3\mathbf{r}\, \text{Im}[\mathbf{E}_a(\mathbf{r}) \cdot \mathbf{P}(\mathbf{r})], \tag{32}$$

where we introduced the auxiliary electric field $\mathbf{E}_a(\mathbf{r})$ equal to the incident field from conjugate sources [cf. Eq. (22)]:

$$\mathbf{E}_a(\mathbf{r}) \stackrel{\text{def}}{=} i\omega\mu_0 \lim_{V_0 \to 0} \int_{V_s \setminus V_0} d^3\mathbf{r}'\, \overline{\mathbf{G}}(\mathbf{r}, \mathbf{r}') \cdot \mathbf{J}_s^*(\mathbf{r}') - i\frac{\mathbf{J}_s^*(\mathbf{r})}{3\omega\varepsilon_h}. \tag{33}$$

To calculate this field, we note that the conjugation of $\mathbf{J}_s^*$ [Eq. (21)] is equivalent to the inversion of sign of $z - z_0$, which leads to the same inversion in Eq. (23), since the sign of $\tilde{z}$ can be freely changed inside the integral in the latter equation. Therefore, analogously to Eq. (25) we obtain

$$\mathbf{E}_a(\mathbf{r}) = -\frac{q\omega}{2\pi\varepsilon_0 m_h^2 v^2 \gamma_h} \exp\left[-i\frac{\omega}{v}(z-z_0)\right] \begin{pmatrix} \frac{x-x_0}{b} K_1\left(\frac{\omega b}{\gamma_h v}\right) \\ \frac{y-y_0}{b} K_1\left(\frac{\omega b}{\gamma_h v}\right) \\ \frac{i}{\gamma_h} K_0\left(\frac{\omega b}{\gamma_h v}\right) \end{pmatrix}$$

$$= \begin{pmatrix} -E_{\text{inc},x}(\mathbf{r}) \\ -E_{\text{inc},y}(\mathbf{r}) \\ E_{\text{inc},z}(\mathbf{r}) \end{pmatrix} \exp\left[-2i\frac{\omega}{v}(z-z_0)\right],$$

(34)

where the additional minus arises from each derivative with respect to $z$.

Equation (32) is very convenient since the integration is performed over the volume of the particle (rather than that of sources), $\mathbf{E}_a(\mathbf{r})$ is easily obtained from the known $\mathbf{E}_{\text{inc}}(\mathbf{r})$, and $\mathbf{P}(\mathbf{r})$ is efficiently calculated in the DDA. Moreover, the resulting expression (32) for $W_{\text{enh}}$ is very similar to the expression (17) for $W_{\text{ext}}$ and can be calculated in a similar way.

Previously, the particle-induced energy losses, expressed as an integral over the volume of the particle, were known only for the specific case of vacuum as a host medium,[8,9,36] for which it is known that $W_{\text{enh}} = W_{\text{ext}}$. In our general approach the equality of Eqs. (32) and (17) follows from the fact that $\mathbf{E}_a(\mathbf{r}) = -\mathbf{E}_{\text{inc}}^*(\mathbf{r})$ which is true if and only if $\gamma_h \in \mathbb{R}$ (i.e., when $\beta_h < 1$). Alternatively, the same can be obtained from the total energy budget (Fig. 1) using $W_0 = 0$ and $W_{\text{rad}} = W_{\text{sca}}$, where the latter follows from the possibility of extending the integration surface for $W_{\text{sca}}$ to infinity and the exponential decay of $\mathbf{E}_{\text{inc}}(\mathbf{r})$. The fast decay of $\mathbf{E}_{\text{inc}}(\mathbf{r})$ is actually required for any excitation that has $W_0 = 0$, since $W_0$ can also be computed as the far-field integral.

The major novelty of Eq. (32) is its applicability to the arbitrary passive host medium ($m_h \in \mathbb{C}$) and the Cherenkov case ($\beta_h > 1$), when $W_{\text{enh}} \neq W_{\text{ext}}$. Although in this case $W_0$ is not zero and may even be effectively infinite (see Section 2.3), the expression for (additional) particle-induced energy losses is well defined. However, the equality $\mathbf{E}_a(\mathbf{r}) = -\mathbf{E}_{\text{inc}}^*(\mathbf{r})$ also remains approximately valid in the quasi-static case [cf. Eq. (S15)], especially when $\beta_h \approx 1$. Thus, to notice new effects, predicted by the rigorous theory of this section for the Cherenkov case, one needs to consider relatively large particles and $\beta_h$ significantly larger than 1.

## 2.5 Electron energy-loss probability

Total energy loss for the electron will be the sum of free-space and particle-induced energy losses. To find it we apply Eq. (S6) to $W_{\text{enh}}$ and $W_0$

$$\Delta E = \Delta E_0 + \Delta E_{\text{enh}} = \frac{2}{\pi}\int_0^\infty d\omega(W_0 + W_{\text{enh}}) = \int_0^\infty d(\hbar\omega)\Gamma_{\text{EELS}}(\hbar\omega)\hbar\omega, \tag{35}$$

where we introduced the electron energy-loss probability density function

$$\Gamma_{\text{EELS}}(\hbar\omega) \stackrel{\text{def}}{=} \frac{2}{\pi}\frac{W_0 + W_{\text{enh}}}{\hbar^2\omega} \tag{36}$$

of the random variable $\hbar\omega$ such that $\Delta E$ is the expected value of this variable.

Strictly speaking, the purely classical theory is deterministic, i.e. each electron should lose exactly $\Delta E$ given by Eq. (35). By contrast, the quantum description of EELS predicts that each electron loses energy in quanta (0 or some natural number of them), each of them is chosen randomly from some probability distribution. However, it has been shown[37–39] that the phenomenological introduction of probability density by Eq. (36) allows the classical theory to correctly reproduce the quantum result in the single-loss (weak-coupling) regime, i.e. when at most one energy quanta is assumed to be lost (also called the Born approximation). This regime is commonly satisfied in EELS experiments, although the strong-coupling regime is gaining increasing interest as well.[40]

The total (measurable) loss probability $\Gamma_{\text{EELS}}$ is proportional to the sum of $W_0$ and $W_{\text{enh}}$ and is naturally the simplest, when $W_0 = 0$ (i.e. $\beta_h < 1$). Otherwise, there are several potential issues. First one is related to the definition of $W_0$, especially for the absorbing host medium (see Section 2.3). Second, any non-zero $W_0$ has a finite value per unit length, thus becomes effectively infinite for unbounded host media (as assumed in the theoretical derivation). In practical applications the thickness of the host medium can be much larger than the particle size, but still finite. In this case, one can hope that Eq. (32) remains approximately correct, while $W_0$ can be computed from Eq. (30) (or another expression for absorbing host medium)[33] with possible addition of the transition-radiation losses.[41] Third, the resulting $W_0$ need to be sufficiently small for the weak-coupling regime to remain valid. The latter requirement can probably be relaxed, assuming that an electron loses at most one energy quantum by interaction with particle, but potentially many quanta by interaction with bulk host medium (and these two processes are independent). However, in such case the loss probabilities proportional to $W_0$ and $W_{\text{enh}}$ cannot be added, but rather need to be convoluted as functions of $\omega$ (each including the zero-loss peak for normalization). But we are not aware of existing rigorous analysis of such option.

In any case, $W_0$ is fully determined by the outer boundaries of the host medium (with respect to the electron beam) and is independent from the particle. By contrast, the focus of this paper is on calculating the particle-induced energy losses $W_{enh}$, for which one generally needs to employ the DDA or other numerical method. To avoid confusion, we define the corresponding particle-induced loss probability as $P_{EELS}$ and aim to compute only this part. When $W_0 = 0$, it is exactly the measurable quantity $\Gamma_{EELS}$, otherwise it is a first step in obtaining $\Gamma_{EELS}$ with the second step consisting of a separate calculation of $W_0$. In the case of a weak-coupling regime for the whole system (particle + slab of host medium) $P_{EELS}$ is equal to the difference between the losses in this system and the losses in the same system without a particle (both of which are potentially measurable). The same approach is commonly used, e.g., when defining cross sections for light scattering by particles in absorbing host medium.[42,43]

The explicit expression for $P_{EELS}$ follows from Eq. (36):

$$P_{EELS}(\hbar\omega) \stackrel{\text{def}}{=} \frac{2}{\pi}\frac{W_{enh}}{\hbar^2\omega} = \frac{m_h'\varepsilon_0 E_0^2}{\pi\hbar^2 k_0} C_{enh}, \tag{37}$$

where we defined the enhancement cross-section

$$C_{enh} \stackrel{\text{def}}{=} -\frac{k_0}{m_h'\varepsilon_0 E_0^2}\int_{V_{int}} d^3\mathbf{r}' \text{Im}[\mathbf{E}_a(\mathbf{r}') \cdot \mathbf{P}(\mathbf{r}')], \tag{38}$$

analogously to $C_{ext}$ [cf. Eqs. (17), (19)]. In the case of $\beta_h < 1$ we have $C_{enh} = C_{ext}$ and Eqs. (37), (38) simplify to a previously known expression for the electron energy-loss probability.[8] Usually, energy losses $\delta E = \hbar\omega$ are expressed in the units of eV, then $P_{EELS}$ is in the units of eV$^{-1}$.

## 2.6 Cathodoluminescence probability

As discussed in Section 2.4, in vacuum (or more generally when $\beta_h < 1$) the radiation from the particle-electron system $W_{rad}$ is only the one produced by induced currents in the particle ($W_{sca}$). In arbitrary non-absorbing host medium, the radiated energy is the energy lost by the electron minus the energy absorbed by the particle:

$$\Delta E = \Delta E_0 + \Delta E_{enh} - \Delta E_{abs} = \frac{2}{\pi}\int_0^\infty d\omega(W_0 + W_{enh} - W_{abs})$$
$$= \int_0^\infty d(\hbar\omega)\Gamma_{CL}^{tot}(\hbar\omega)\hbar\omega, \tag{39}$$

where by analogy to $\Gamma_{EELS}(\hbar\omega)$ we introduce the total light-emission probability density

$$\Gamma_{CL}^{tot}(\hbar\omega) \stackrel{\text{def}}{=} \frac{2}{\pi}\frac{W_0 + W_{enh} - W_{abs}}{\hbar^2\omega}. \tag{40}$$

As discussed in Section 2.5, $W_0$ is potentially problematic, but the computationally-intensive method is only needed to calculate the particle-related contribution to $\Gamma_{\text{CL}}^{\text{tot}}$, which we further denote as $P_{\text{CL}}^{\text{tot}}$:

$$P_{\text{CL}}^{\text{tot}}(\hbar\omega) \stackrel{\text{def}}{=} \frac{2}{\pi}\frac{W_{\text{rad}} - W_0}{\hbar^2\omega} = \frac{m_{\text{h}}'\varepsilon_0 E_0^2}{\pi\hbar^2 k_0}(C_{\text{enh}} - C_{\text{abs}}) = \frac{m_{\text{h}}'\varepsilon_0 E_0^2}{\pi\hbar^2 k_0}(C_{\text{enh}} - C_{\text{ext}} + C_{\text{sca}}), \quad (41)$$

where the second part prevents the loss of precision when $C_{\text{enh}}$ and $C_{\text{abs}}$ are equal within a few decimal places, which is often the case for metallic nanoparticles. In the latter case one may calculate $C_{\text{sca}}$ by integrating the far field instead of using the relation $C_{\text{sca}} = C_{\text{ext}} - C_{\text{abs}}$.

The above definition of $P_{\text{CL}}^{\text{tot}}$ is fundamentally the most natural, since it accounts for the total outgoing radiation in all directions (denoted by superscript "tot"). However, all existing measurement modalities for CL collect light only in the upper hemisphere when the electron moves downwards, this always excludes the Cherenkov cone (directed in the lower hemisphere). To be more accurate, one should consider a large but finite chunk of the host medium (as in Section 2.5) and draw the Cherenkov cone from each point of the electron trajectory inside the host medium. First, this solves the problem with potential unboundedness of $W_0$, since it does not contribute to the measured signal if we neglect the reflection of the Cherenkov radiation from the bottom boundary of the host medium. Second, the interference of $\mathbf{E}_{\text{inc}}$ and $\mathbf{E}_{\text{sca}}$ in the far-field (also concentrated in the direction of the Cherenkov cone) is irrelevant to the measurements as well. But this interference is exactly the one which leads to the difference between $W_{\text{ext}}$ and $W_{\text{enh}}$. Thus, to exactly reproduce the experimental signal one should integrate $|\mathbf{E}_{\text{sca}}|^2$ over the detector collection angle, which is often not known. However, if one aims to have a simple approximation, the integral of the scattered intensity over the whole solid angle (proportional to $C_{\text{sca}}$) is a reasonable option, at least for particles smaller than the wavelength. For the latter case, the angular dependence of $|\mathbf{E}_{\text{sca}}|^2$ is relatively weak, thus the ratio between the integral over the detector collection angle and $C_{\text{sca}}$ is expected to weakly depend on the loss energy (frequency). Then the simulated spectrum will reproduce the measured one semi-quantitatively up to a constant factor. Therefore, we postulate the CL probability to be:

$$P_{\text{CL}}(\hbar\omega) \stackrel{\text{def}}{=} \frac{2}{\pi}\frac{W_{\text{sca}}}{\hbar^2\omega} = \frac{m_{\text{h}}'\varepsilon_0 E_0^2}{\pi\hbar^2 k_0} C_{\text{sca}}, \quad (42)$$

which agrees with the ones used in other simulation methods (Lorenz-Mie, BEM). Naturally, we also have $P_{\text{CL}} = P_{\text{CL}}^{\text{tot}}$ for $\beta_{\text{h}} < 1$. Finally, we stress once again that in the case of $\beta_{\text{h}} > 1$ the difference between $W_\text{ext}$ and $W_\text{enh}$ is important for the electron energy losses (Section 2.5), but is not visible in existing CL measurement configurations.

The definition of $P_\text{CL}$ becomes even more problematic in the case of absorbing host medium, since the absorption of the scattered radiation is generally significant. To accurately reproduce the signal at a detector one should account for variation of optical path in the host medium with the scattering angle. The corresponding attenuation will depend on variation of $\text{Im}\, m_\text{h}$ with $\omega$. Still, we can expect Eq. (42) to qualitatively reproduce the measured CL spectra. Fortunately, the far limit of the incident electron field, which is not negligible for $\text{Re}\,\beta_\text{h} > 1$, has the same behavior as attenuated Cherenkov radiation. Thus, its contribution to detector signal can also be usually neglected. By contrast, $P_\text{CL}^\text{tot}$ has no practical relevance at all in this case, since all underlying components of the energy balance depend on the outer boundary of the host-medium and the far-field powers cannot be computed as integrals over the particle volume.

Finally, we stress that neither $P_\text{EELS}$ nor $P_\text{CL}^\text{tot}$ are generally guaranteed to be positive. While the relations $W_\text{abs} \geq 0$ and $W_\text{sca} > 0$ (hence, $P_\text{CL} > 0$) always hold, see Eq. (18) and [23], only in the case $\beta_\text{h} < 1$ the probabilities $P_\text{EELS}$ and $P_\text{CL}$ are proportional to $W_\text{abs} + W_\text{sca}$ and $W_\text{sca}$, respectively, and hence are strictly positive. Negative probabilities, that may be obtained in absorbing or Cherenkov host medium, are analogous to the phenomenon of negative extinction [42]. Note, however, that $W_\text{ext} > 0$ holds in any non-absorbing medium, including the case $\beta_\text{h} > 1$.

## 2.7 Scale invariance

To account for the host medium in the DDA, the Green's tensor, the incident electric field, and the particle refractive index must be modified according to $m_\text{h}$. For the electromagnetic codes that do not natively support refractive index of the host medium as an input parameter, this can be done by scaling other input parameters. For the case of the plane-wave excitation and $m_\text{h} > 0$, it is sufficient to divide both $m_\text{p}$ and the vacuum wavelength $\lambda$ by $m_\text{h}$, resulting in the correct values of $m(\mathbf{r})$ and $k$.[44] All computed values are then correct, except for scaling of cross sections since $I_0$ scales with $m_\text{h}$, when $E_0$ is fixed [Eq. (19)].

In the case of the electron excitation, additional care is required to keep the incident field [Eq. (25)] the same up to a constant factor. Relativistic factors $\beta_\text{h}$ and $\gamma_\text{h}$ need to be the same, since they differently affect the fields along the transverse and longitudinal coordinates, leading to the scaling $v \to m_\text{h} v$. Obviously, this applies only to the case of not very dense host medium ($\beta_\text{h} < 1$) – this derivation additionally illustrates that this case is fully analogous to the case of vacuum. The scaling of $v$ is compensated by scaled $\omega$, corresponding to the above scaling of $\lambda$. The resulting $\mathbf{E}_{\text{inc}}(\mathbf{r})$ in vacuum is then $m_{\text{h}}$ times the correct field in the host medium.

Let us generalize this analysis to arbitrary scaling of refractive index by $\eta \in \mathbb{R}$ combined with scaling of all lengths by $\xi \in \mathbb{R}$ (and inverse scaling of $k$). The latter corresponds to the classical scale-invariance rule in electromagnetic scattering.[45] Formally, we have

$$m_{\text{h}} \to m_{\text{h}}/\eta, \qquad m_{\text{p}} \to m_{\text{p}}/\eta, \qquad v \to \eta v,$$
$$\mathbf{r} \to \mathbf{r}/\xi, \qquad \mathbf{r}_0 \to \mathbf{r}_0/\xi, \qquad k \to \xi k. \tag{43}$$

Which, after straightforward algebraic analysis [Eqs. (7),(17)–(19),(25),(26),(32),(37),(41)], leads to

$$\omega \to \xi\eta\omega, \qquad \mathbf{E}_{\text{inc}} \to \xi\eta\mathbf{E}_{\text{inc}}, \qquad \mathbf{E} \to \xi\eta\mathbf{E}, \qquad \mathbf{P} \to \xi\mathbf{P}/\eta, \qquad \overline{\mathbf{G}} \to \xi\overline{\mathbf{G}},$$
$$I_0 \to I_0/\eta, \qquad \frac{\partial W_0}{\partial z} \to \xi\eta\frac{\partial W_0}{\partial z}, \qquad W_X \to \eta W_X, \qquad C_X \to \eta^2 C_X, \tag{44}$$
$$P_{\text{EELS,CL}} \to P_{\text{EELS,CL}}/\xi, \qquad \Delta E_X \to \xi\eta^2 \Delta E_X.$$

where $X$ is a subscript of any power (enh, ext, abs) and $E_0$, $c$, $\hbar$, $\varepsilon_0$, $\mu_0$ are kept intact. Note that the scaling of powers and cross sections is different from the case of plane-wave excitation due to the additional scaling of the incident field.

Coming back to simulating EELS and CL in the host medium with $\beta_{\text{h}} < 1$ using any vacuum-based code, one has two options. The first one, described above, corresponds to $\eta = m_{\text{h}}(\omega)$ and $\xi = 1$. The particle geometry and electron trajectory are kept intact, only the electron speed, the particle refractive index, and simulation frequency (energy loss or vacuum wavelength) need to be scaled:

$$v \to m_{\text{h}}v, \qquad m_{\text{p}} \to m_{\text{p}}/m_{\text{h}}, \qquad \hbar\omega \to m_{\text{h}}\hbar\omega \ (\lambda \to \lambda/m_{\text{h}}). \tag{45}$$

Note that the values of both $m_{\text{p}}$ and $m_{\text{h}}$ correspond to the original frequency rather than to the scaled one. Thus, both $P_{\text{EELS}}$ and $P_{\text{CL}}$ computed with vacuum code for energy loss $m_{\text{h}}\hbar\omega$ are exactly the sought probabilities for energy loss $\hbar\omega$ in the host medium. The second option keeps $\omega$ and $\lambda$ intact at the cost of additional geometrical scaling, i.e. $\eta = m_{\text{h}} = 1/\xi$. Thus, additionally both particle dimensions and electron position need to be multiplied by $m_{\text{h}}$:

$$v \to m_{\text{h}}v, \qquad m_{\text{p}} \to m_{\text{p}}/m_{\text{h}}, \qquad \mathbf{r} \to m_{\text{h}}\mathbf{r}, \qquad \mathbf{r}_0 \to m_{\text{h}}\mathbf{r}_0,$$
$$P_{\text{EELS,CL}} \to m_{\text{h}}P_{\text{EELS,CL}}. \tag{46}$$

In this case, the probability density computed by the vacuum code needs to be further divided by $m_{\text{h}}$. Finally, note that the scaling $v \to m_{\text{h}}v$ corresponds to the following scaling of electron kinetic energy $\mathcal{E}$:

$$\mathcal{E} \to \mathcal{E}_0 \left( \frac{\mathcal{E} + \mathcal{E}_0}{\sqrt{\mathcal{E}_0^2 - (\varepsilon_h - 1)\mathcal{E}(2\mathcal{E}_0 + \mathcal{E})}} - 1 \right), \tag{47}$$

where $\mathcal{E}_0 = m_e c^2$ is the electron rest mass energy.

Naturally, the proposed scaling cannot transform the Cherenkov case into a vacuum one, but an alternative approximate scaling for the case $\beta_h > 1$ is described in Section S4 of the Supporting Information.

## 3 Simulations

### 3.1 Implementation in ADDA

To perform EELS and CL simulations according to the developed general theory, we modified the open-source software ADDA.[21] A single run of ADDA performs a simulation for a single set of parameters ($\lambda$, beam position, etc.). To simulate a loss spectrum or to scan a particle's cross-section with the beam, it is necessary to run ADDA multiple times varying the desired parameters. For instance, to simulate the EELS and CL spectra one needs to vary $\lambda$ and the corresponding $m_p$. We developed a Python wrapper, named `ADDAwrapper`, to automate this process.

The wrapper is designed to work as a Python library, so one only has to fill the example preset file with the simulation parameters, and call high-level functions from this file to perform the corresponding set of simulations (spectrum, loss probability scan over the cross-section of the particle, etc.), collect the data from these simulations, and plot it in various formats. Ready-to-use examples are distributed with the ADDA in the folder `/examples/`.

`ADDAwrapper` supports multithreading to speed up the simulation up to an order of physical processor cores. Apart from that, the examples use optimized set of simulation parameters, which allows the wrapper to perform EELS and CL simulations with ADDA up to an order of magnitude faster than with the default code settings. More details are given in Section S5 of the Supporting Information.

### 3.2 Comparison to the Lorenz-Mie theory in vacuum

The classical Lorenz-Mie theory presents a solution of light scattering by a homogeneous sphere.[46] This solution was extended for calculation of the EELS and CL by Garcia de Abajo et al.,[5,47] who also implemented it in the online simulation tool,[48] which we used to obtain the reference solutions.

In Fig. 2. we present the EELS and CL spectra simulated with ADDA and the Lorenz-Mie theory. In vacuum an electron with kinetic energy $\mathcal{E} = 100$ keV passes at a distance of 100 nm from the center of a silver 75-nm-radius sphere. Optical data for silver is taken from [49], since it is available as a built-in option in the Lorenz-Mie simulation tool, and 128 dipoles along the $x$-axis ($n_x = 128$) are used for volume discretization. Step size of 0.05 eV is used for all simulations of spectra.

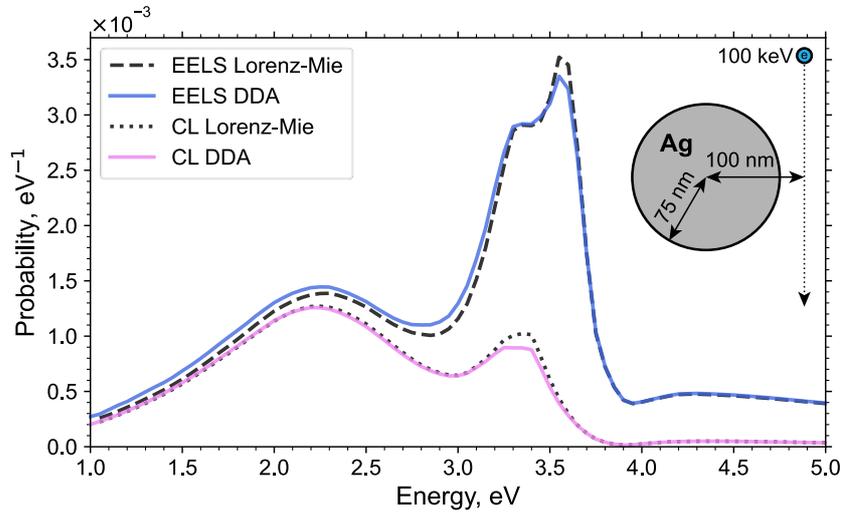

Fig. 2. EELS and CL spectra of a sphere simulated with the Lorenz-Mie theory and ADDA ($n_x = 128$). The problem parameters are shown in the inset.

Simulated spectra are close to the exact solution, but can be further improved by increasing $n_x$ (refining discretization) at the expense of extra computational resources. A more efficient approach is to employ the Richardson extrapolation.[50] Although it is a semi-empirical method, it was successfully used in various applications[51–54] and was especially efficient for nanoparticles. In particular, we used the guidelines from [50], performing simulations for a set of $n_x = 128, 108, 91, 76, 64, 54, 45, 38$, and 32 and extrapolating the dependence on the dipole size to $d = 0$ using the quadratic function (see Fig. 3 for an example). The errors of data points were assumed to scale as $d^3$, and the standard error of fitted value at $d = 0$ was multiplied by 2 to obtain nominal 95% confidence interval.

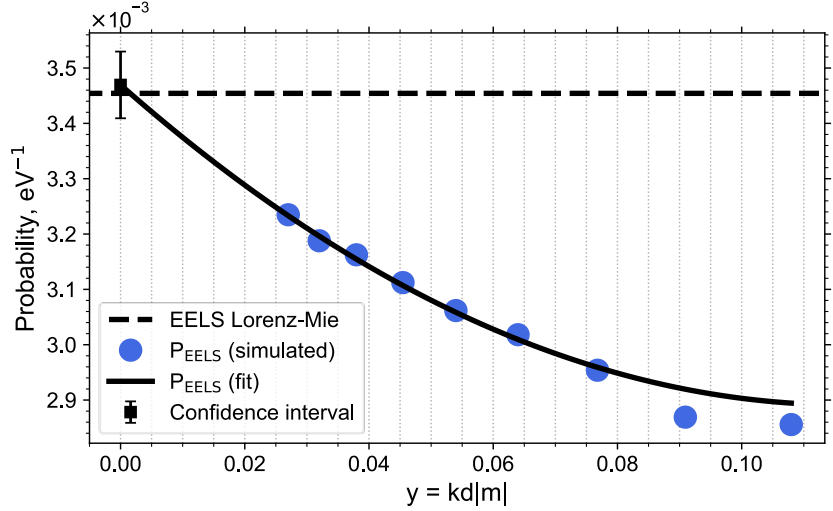

Fig. 3. Extrapolation of the simulated $P_{\text{EELS}}$ value at $\delta E = 3.6$ eV. The parameters of the problem are the same as in Fig. 2.

In Fig. 4 the same Richardson extrapolation applied to the whole spectrum leads to almost perfect match with the exact solution. The extrapolation is performed automatically by ADDAwrapper both for a single $\delta E$ and for the whole spectrum.

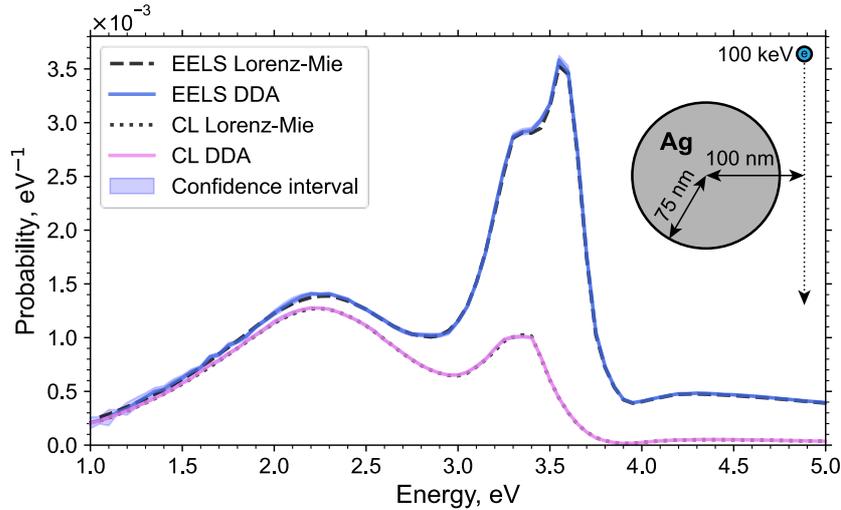

Fig. 4. EELS and CL spectra simulated with the Lorenz-Mie theory and ADDA ($n_x = 128$ with the extrapolation). 95% confidence interval is showed for the extrapolated values. The parameters of the problem are shown in the inset (the same as in Fig. 2).

## 3.3 Comparison to the BEM in a host medium

Next, we consider particles embedded in a host medium. In Fig. 5 we show how EELS spectrum changes when the same sphere is placed inside a non-absorbing host medium, in comparison to the simulation for vacuum ($m_{\text{h}} = 1$). In host media the peaks shift to the lower energies and their magnitudes decrease. We also performed scaling of separate DDA simulations for a particle in vacuum according to Section 2.7 – both approaches there [Eqs. (45), (46)] lead to identical results (only one of them is shown). For $m_\mathrm{h} = 1.5$ the scaled spectrum perfectly matches the one obtained by directly setting this value of $m_\mathrm{h}$ in the code (Fig. 5). By contrast, for $m_\mathrm{h} = 2$ the electron with 100 keV kinetic energy moves faster ($0.55c$) than the speed of light in this medium ($0.5c$), causing qualitative differences in theoretical description of this process ($\beta_\mathrm{h} = 1.10$). Specifically, the exact scaling is not valid anymore, but we have tried an approximate correction given by Eq. (S16) in the Supporting Information. One can see that such scaling qualitatively reproduces the data, both peak positions and, less accurately, amplitudes. However, it is less accurate for peak magnitudes of CL (Fig. S1). See also Fig. S2 for further discussion of this scaling as a function of $m_\mathrm{h}$.

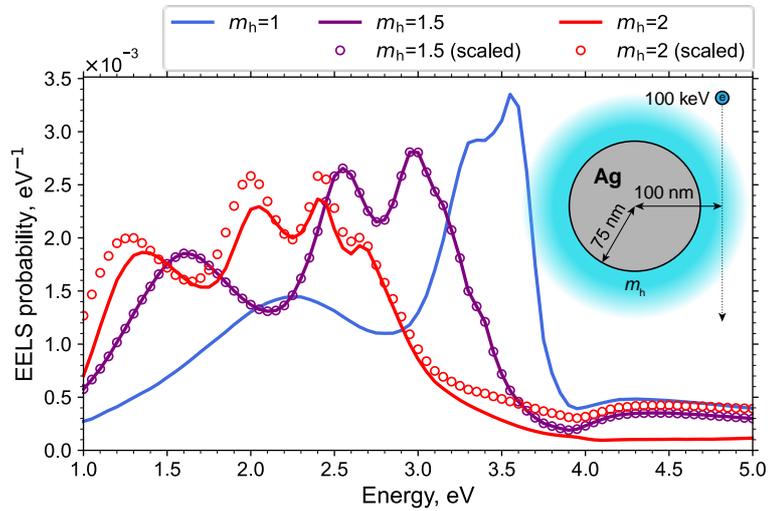

Fig. 5. EELS spectra for a sphere in an infinite non-absorbing host medium simulated with ADDA ($n_x = 128$), either directly or through scaling of simulations in vacuum (see the text for details). The problem parameters are shown in the inset. In the host medium with $m_\mathrm{h} = 2$ the electron is faster than light, causing the Cherenkov radiation.

We further compare our results with the BEM, as implemented in the MNPBEM17 code,[10] as it seems to support arbitrary host medium. We limit ourselves to EELS in this section, presenting the corresponding CL results in the Supporting Information (Fig. S3–Fig. S5). First, we reproduce the EELS data from Fig. 4. As shown in Fig. 6, the BEM agrees with the DDA and the Lorenz-Mie theory except for the lowest loss energies (<1.4 eV). We used 1024 surface points for all BEM simulations, and further mesh refinement did not improve the accuracy for such energies. We hypothesize that it is related to large $|m_\mathrm{p}| > 7$ in this range, but it is not important for further discussion. Note also that CL results for BEM do not have this artifact (Fig. S3). Since the online tool[48] does not allow manual specification of a dielectric function (which is used below), we further use the Lorenz-Mie theory built into MNPBEM. However, we tested that they produce identical results for the case of Fig. 6 (data not shown).

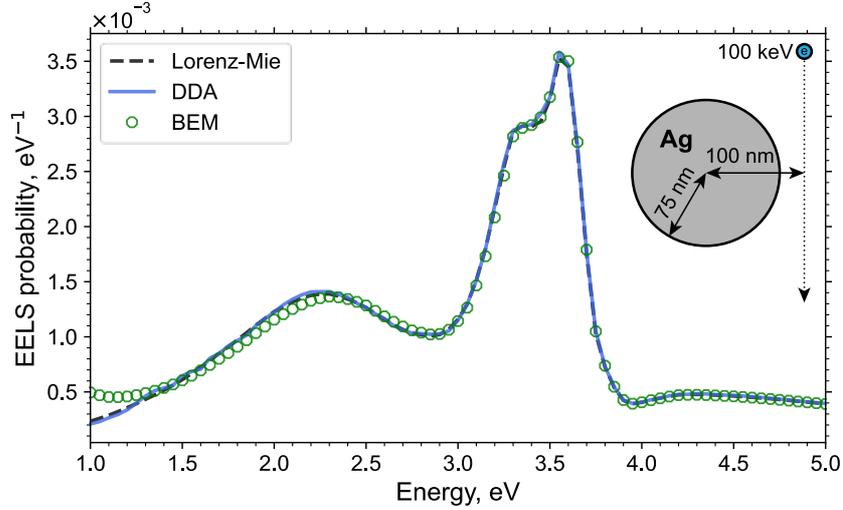

Fig. 6. EELS spectra simulated with the Lorenz-Mie theory, ADDA ($n_x = 128$ with the extrapolation), and the BEM. The parameters of the problem are shown in the inset (the same as in Fig. 4).

Next, we set $m_h = 1.5$ in both MNPBEM and ADDA and compare the results with the Lorenz-Mie theory, scaled according to Eq. (45). The results in Fig. 7 show that all three methods agree. The visible inaccuracy of the DDA at the peak value at 3.0 eV disappears with further grid refinement to $n_x = 192$ with the extrapolation (data not shown). The CL results also agree, but only if the MNPBEM results are additionally multiplied by $m_h$ (Fig. S4).

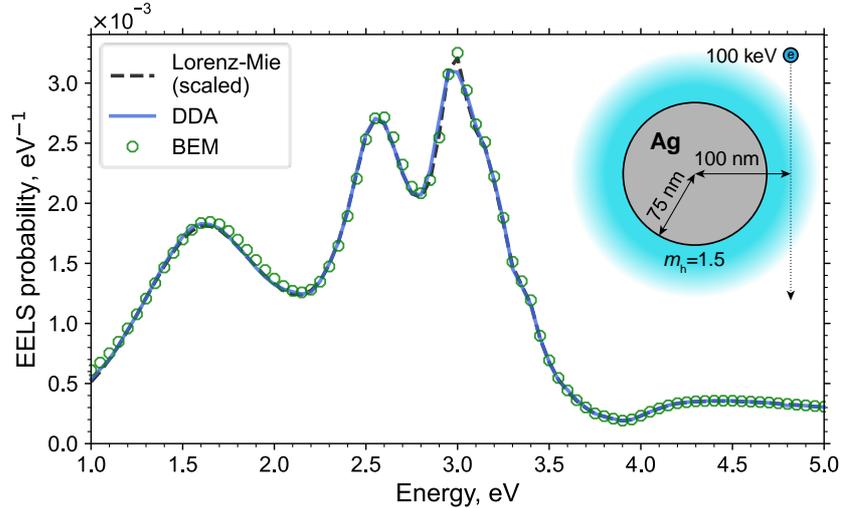

Fig. 7. EELS spectra simulated with the Lorenz-Mie theory (using scaling), ADDA ($n_x = 128$ with the extrapolation), and BEM. The parameters of the problem are shown in the inset (subset of Fig. 5).

The situation is markedly different for $m_h = 2$ (Cherenkov case), as shown in Fig. 8 featuring huge difference between the DDA and BEM. To investigate this issue, let us formally define the extinction probability $P_{\text{ext}}$ by analogy to $P_{\text{EELS}}$ [Eq. (37)] as

$$P_{\text{ext}}(\hbar\omega) \stackrel{\text{def}}{=} \frac{2}{\pi}\frac{W_{\text{ext}}}{\hbar^2\omega} = \frac{m_h' \varepsilon_0 E_0^2}{\pi \hbar^2 k_0} C_{\text{ext}}. \tag{48}$$

As discussed in Section 2.5, $P_{\text{ext}}$ is identical to $P_{\text{EELS}}$ for $\beta_h < 1$, which we explicitly tested for the case of $m_h = 1.5$ (data not shown). However, they are different for $m_h = 2$, as shown in Fig. 8. And, surprisingly, the spectrum simulated with BEM turns out to be matching $P_{\text{ext}}$ values instead of $P_{\text{EELS}}$. Thus, we conclude that MNPBEM works fine for $\beta_h < 1$, but is incorrect for the Cherenkov case. It uses the correct incident field in the latter case, but seems to use incorrect expression for $P_{\text{EELS}}$ based on equivalence of $W_{\text{ext}}$ and $W_{\text{enh}}$ that no longer holds. The latter equivalence is also implicitly used in other DDA codes for EELS simulations: $e$-DDA[9] and DDEELS.[8] By contrast, the results of $P_{\text{CL}}$ computed with MNPBEM for $m_h = 2$ agree with the DDA after multiplication by $m_h$ (Fig. S5). This is expected, since the definition of $P_{\text{CL}}$ is not affected by the difference between $W_{\text{ext}}$ and $W_{\text{enh}}$ (in contrast to $P_{\text{CL}}^{\text{tot}}$), as discussed in Section 2.6.

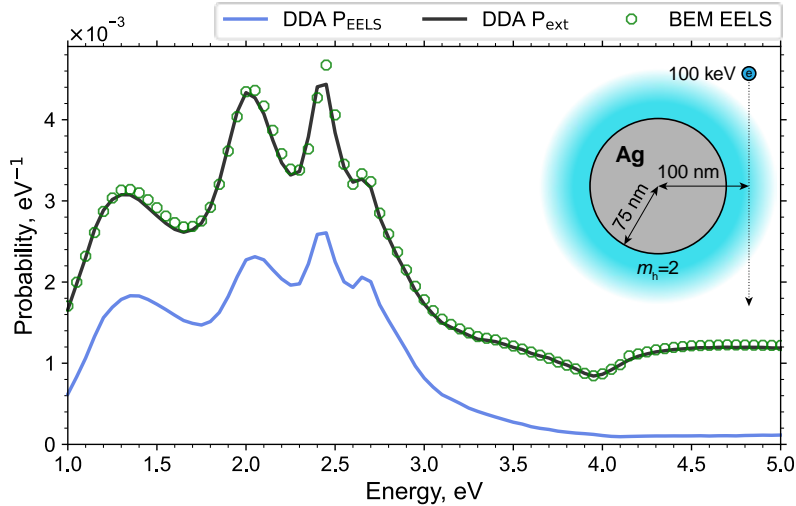

Fig. 8. EELS and extinction probabilities simulated with the DDA ($n_x = 128$ with the extrapolation) compared to EELS spectrum simulated with BEM. The parameters of the problem are shown in the inset (subset of Fig. 5).

### 3.4 Comparison to the experiments

In EELS experiments a particle is usually placed on or inside a substrate. Raza et al.[11] investigated silver nanospheres encapsulated in silicon nitride, which fixes them in place and prevents silver oxidation. In that paper both experimental and simulated spectra are shown for an encapsulated nanosphere with a radius of 9.2 nm and an electron passing at different impact parameters. The simulations were performed for a full system geometry, discretizing the finite chunk of the silicon nitride layer.

By contrast, we try to reproduce these experimental data by simulations for a particle in an infinite host medium. Specifically, Fig. 9 shows the simulated spectrum for the same silver nanosphere along with the experimental data from [11] for an impact parameter of 12.4 nm. The simulation is done in ADDAwrapper for $m_h = \sqrt{3.2}$, $\mathcal{E} = 100$ keV, and $n_x = 128$. This is a Cherenkov medium with $\beta_h = 1.05$. The optical data for silver is taken from [55], which is somewhat more accurate than the data used in theoretical comparisons of Sections 3.2 and 3.3. The simulated spectrum was additionally convoluted with Gaussian point spread function (PSF) with FWHM of 0.15 eV to match the experimental conditions. The simulated spectrum matches the main 2.8 eV peak position, as well as its right shoulder. The remaining disagreement at the left side of the peak can potentially be related to imperfect zero-loss peak removal, which also causes some experimental values at larger loss energies to systematically lie below zero. While such negative values are not impossible when $\beta_h > 1$ (see Section 2.6), they are not observed in simulations for the specific problem parameters.

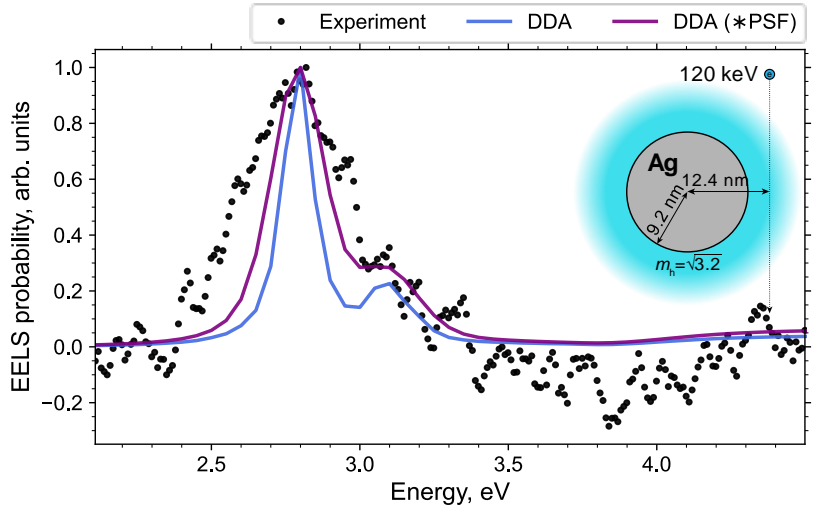

Fig. 9. Experimental and simulated EELS spectra for a silver sphere encapsulated in silicon nitride, normalized by the maximum value. The problem parameters are in the inset. Experimental data is taken from [11], simulation is done with ADDA using $n_x = 128$ (blue line) and convoluted with Gaussian PSF with FWHM of 0.15 eV (purple line).

In another work Kobylko et al.[12] investigated gold nanoparticles placed inside a glass substrate. One of them was a 92.6-nm-long gold nanowire with a 7.8 nm radius, investigated by electrons with $\mathcal{E} = 100$ keV. We modeled this nanowire as a perfect cylinder with hemispheres at both ends with a computational grid of 16x16x190 dipoles and placed it in an infinite medium with $m_\mathrm{h} = 1.45$ (corresponding to $\beta_\mathrm{h} = 0.79$); optical data for gold is taken from [55]. We compare the results of the DDA simulations with the experimental data from [12] in Fig. 10. The experimental data include EELS spectrum averaged over the cross section of the nanowire (Fig. 10, a), as well as plasmon maps for four observed resonant energies (Fig. 10, b–e): 0.86, 1.27, 1.66, and 2.4 eV. The simulation results include the same cross-section-averaged spectrum (Fig. 10, f) and the plasmon maps for resonant energies of the simulated spectrum (Fig. 10, g–j): 0.7, 1.2, 1.55, and 2.4 eV, which are slightly different from the experimental ones. The latter is probably caused by imperfect cylindrical geometry of the particle in the experiment. Apart from that and expected broadening of the experimental peaks, simulations reproduce the experimental data both for EELS spectrum and plasmon maps.

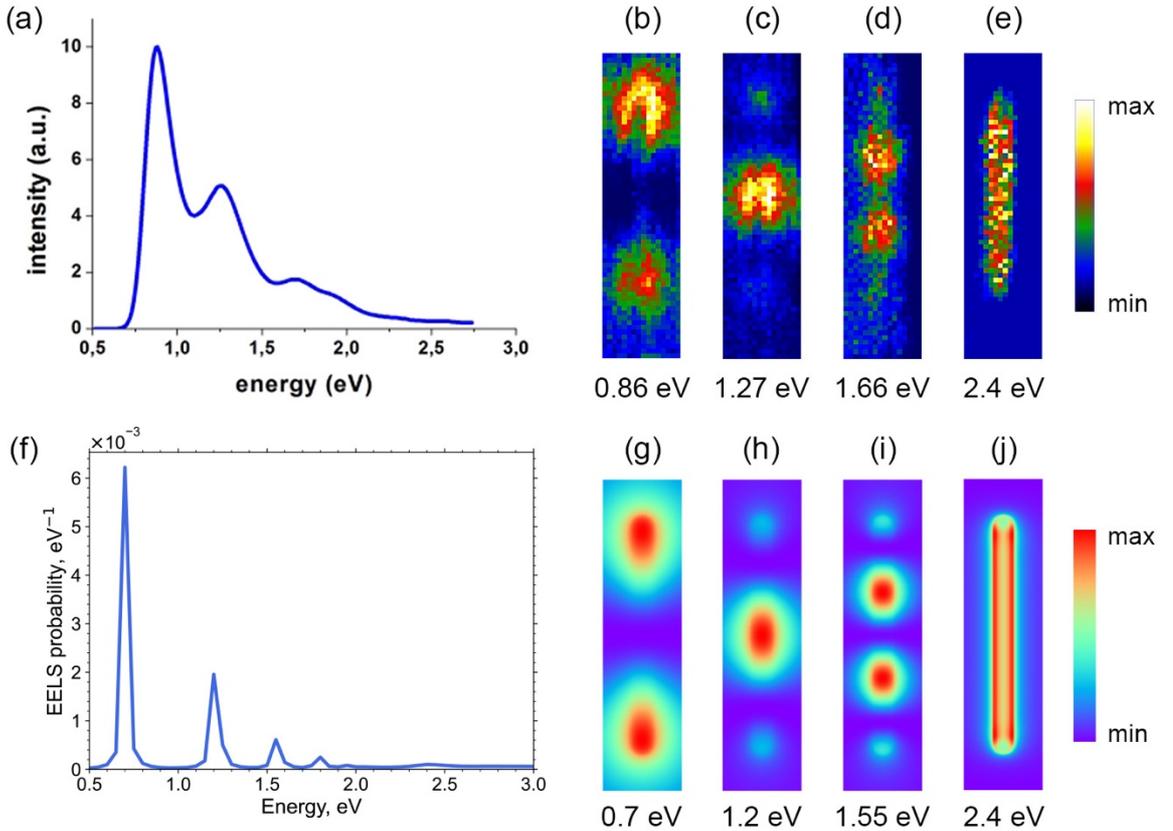

Fig. 10. Comparison of experimental EELS data (a-e) for a gold nanowire in a glass substrate from [12] with the DDA simulations (f-j) (see the text for details). Shown are the EELS spectrum averaged over the cross section of the nanowire (a,f) and plasmon maps for four resonant energies (b-e, g-j). The resonant energies slightly differ between experiments and simulations.

## 4 Conclusion

We derived all the quantities necessary to simulate EELS and CL in the DDA in the energy-budget framework based on the volume-integral equation. This framework has shown its versatility in calculating the incident electric field, free-space energy losses, and particle-induced energy losses of a current source, as well as scattered power corresponding to CL. The obtained expressions either coincide with those known from the literature (derived by several other methods) or appear for the first time. In particular, a volume-integral expression for particle-induced energy losses of arbitrary source [Eq. (32)] can be efficiently calculated in the DDA. This expression remains valid for an arbitrary (even absorbing or Cherenkov) host medium, and in the case of vacuum it reduces to the known one.[8,9,36] We also extended the scale invariance rule to EELS, allowing one to apply any existing vacuum-based code to the case of moderately dense non-absorbing host medium. The presented framework for a homogeneous medium is derived using the free-space tensor $\overline{\mathbf{G}}$, but it can be further generalized to the cases of layered media, such as a semi-infinite or thin substrate, by employing the corresponding Green's tensors.

For a test case of a sphere in vacuum, the DDA results (with extrapolation of the computed values versus the dipole size) almost match the exact Lorenz-Mie solution, exceeding the accuracy of the BEM. The same agreement between the three methods holds for a host medium with no intrinsic electron losses, but the results of the Lorenz-Mie theory need to be scaled (as mentioned above), since only vacuum implementations of this theory are available. For the Cherenkov (sufficiently dense) host medium, the DDA results for EELS probability disagreed with that of BEM (as implemented in the MNPBEM code), since the latter code ignores the interference of the Cherenkov and particle-induced radiations (effectively assuming that EELS probability is proportional to the extinction power). The CL spectra agree between the DDA and BEM in all considered cases up to the overall scale of MNPBEM results (it needed to be additionally multiplied by $m_\mathrm{h}$).

Rigorously accounting for the host medium is important for the corresponding experimental conditions, i.e. when particles or voids are considered in larger homogeneous slabs, since LSPRs strongly depend on dielectric properties of this medium. The implementation of the developed theory in the open-source ADDA code, augmented with a comprehensive Python wrapper, makes any such simulations easily available to practitioners. To illustrate these novel capabilities, we simulated actual EELS experiments: for nanospheres encapsulated in silicon nitride and nanowires inside glass. For the former experiment, the electrons were faster than the speed of light in the substrate (Cherenkov case), and the DDA perfectly reproduced the EELS spectrum in terms of a peak position. For the nanorod inside glass the DDA reproduced the EELS spectrum, as well as the plasmon maps for all four plasmon resonances determined in the experiment.

# 5  Acknowledgments

We thank Søren Raza for providing experimental data shown in Fig. 9 and Mathias Kobylko for providing experimental images shown in Fig. 10. We also thank Tomas Ostaševičius for making the first prototype of EELS simulations in the ADDA code and Alexander Moskalensky for implementing the support of host medium in this code (for light-scattering simulations). The theoretical developments and implementation of EELS simulations in ADDA were supported by Russian Foundation for Basic Research (grant no. 18-01-00502). While the comparisons with BEM and experiments, the study of simulation accuracy, and adding the support of absorbing host medium to ADDA were supported by the Russian Science Foundation (grant no. 18-12-00052).

# Supporting Information for:
# Simulating electron energy-loss spectroscopy and cathodoluminescence for particles in arbitrary host medium using the discrete dipole approximation

Alexander A. Kichigin[1,2] and Maxim A. Yurkin[1,2,*]

[1] *Voevodsky Institute of Chemical Kinetics and Combustion SB RAS,*
*Institutskaya Str. 3, 630090, Novosibirsk, Russia*

[2] *Novosibirsk State University, Pirogova Str. 2, 630090, Novosibirsk, Russia*

*\*Corresponding author: yurkin@gmail.com*

## S1 Fourier transform

We use the following definitions for the direct and inverse Fourier transforms:

$$f(\omega) = \int_{-\infty}^{\infty} \mathrm{d}t f(t) e^{-\mathrm{i}\omega t},$$
$$f(t) = \frac{1}{2\pi} \int_{-\infty}^{\infty} \mathrm{d}\omega f(\omega) e^{\mathrm{i}\omega t}. \quad (S1)$$

Let $f(t)$ and $g(t)$ be real-valued functions, then, according to the Plancherel theorem,

$$\int_{-\infty}^{\infty} \mathrm{d}t f(t) g(t) = \frac{1}{2\pi} \int_{-\infty}^{\infty} \mathrm{d}\omega f(\omega) g^*(\omega). \quad (S2)$$

The real-valuedness of $f(t)$ implies that $f(-\omega) = f^*(\omega)$ and analogously for $g$. Therefore, we split the integration interval and, using substitution $\omega \to -\omega$, re-write Eq. (S2) as

$$\int_{-\infty}^{\infty} \mathrm{d}t f(t) g(t) = \frac{1}{\pi} \int_{0}^{\infty} \mathrm{d}\omega \mathrm{Re}[f(\omega) g^*(\omega)]. \quad (S3)$$

## S2 Energy transfer for sources with general time dependence

In the energy budget framework the expressions for the powers are derived for the case of time-harmonic fields. Any other fields can be represented as a linear superposition of time-harmonic fields, the formal mechanism for this representation is the Fourier transform (see Section S1). First, note that the Fourier transforms of fields, $\mathbf{E}(\mathbf{r})$ and $\mathbf{H}(\mathbf{r})$, have the units of electric and magnetic field *per frequency*, i.e., V·s/m and A·s/m, respectively. As a result, the units of their powers are W·s$^2$=J·s.

We further derive the total energy lost by the electron, starting from the time domain and converting to the frequency one according to Section 1.7.5 of [56]. Let $\mathbf{E}(\mathbf{r}, t)$ and $\mathbf{H}(\mathbf{r}, t)$ be the electric and the magnetic fields respectively, then the time-domain Poynting vector is defined as

$$\mathbf{S}(\mathbf{r}, t) \stackrel{\text{def}}{=} \mathbf{E}(\mathbf{r}, t) \times \mathbf{H}(\mathbf{r}, t), \tag{S4}$$

and the total energy gained or lost inside a closed surface $A$ is

$$\Delta E = \int_{-\infty}^{\infty} dt \oint_A d\mathbf{A} \cdot \mathbf{S}(\mathbf{r}, t). \tag{S5}$$

We use the same symbols for time- and frequency-domain counterparts, but the time argument is always explicitly included for the former to avoid confusion. We further change the order of integration and apply Eq. (S3) to transit to the frequency domain:

$$\Delta E = \oint_A d\mathbf{A} \cdot \int_{-\infty}^{\infty} dt [\mathbf{E}(\mathbf{r}, t) \times \mathbf{H}(\mathbf{r}, t)] = \frac{2}{\pi} \int_0^{\infty} d\omega \oint_A d\mathbf{A} \cdot \frac{1}{2} \text{Re}[\mathbf{E}(\mathbf{r}) \times \mathbf{H}^*(\mathbf{r})]$$
$$= \frac{2}{\pi} \int_0^{\infty} d\omega \, W, \tag{S6}$$

where $W$ is the power for a time-harmonic field [cf. Eqs. (9), (10)]. Eq. (S6) is the universal way to convert time-harmonic powers into the total energy transfers for fields with general time dependence, such as the one for a moving electron.

## S3 Electric field derivation

To calculate the integral defined by Eq. (23), we rewrite it as

$$I_1(\mathbf{r}) = \frac{q}{4\pi} \exp\left[i\frac{\omega}{v}(z - z_0)\right] \int_{-\infty}^{\infty} d\tilde{z} \frac{\exp\left(ik\sqrt{b^2 + \tilde{z}^2} - i\frac{\omega}{v}\tilde{z}\right)}{\sqrt{b^2 + \tilde{z}^2}}, \tag{S7}$$

where $\tilde{z} \stackrel{\text{def}}{=} z - z'$. To calculate the remaining integral, denoted as $I_S(b)$, let us introduce $u \in \mathbb{R}$ such that $\sinh u = \tilde{z}/b$ and use the identity $\cosh^2 u - \sinh^2 u = 1$:

$$I_S(b) = \int_{-\infty}^{\infty} du \exp\left[i\frac{\omega b}{v}(\beta_h \cosh u - \sinh u)\right]. \tag{S8}$$

In the case $\text{Im}\,\beta_h > 0$ we use the integral representation for the Hankel function of the first kind $H_0^{(1)}$ (Eq. 10.9.15 of [32]) by substituting $z = \omega b \beta_h/v$ ($\text{Im}\,z > 0$) and $\zeta = -\omega b/v \in \mathbb{R}$, and then express it as a modified Bessel function $K_0$ (Eq. 10.27.8 of [32]):

$$I_S(b) = \pi i H_0^{(1)}\left(\frac{\omega b}{v}\sqrt{\beta_h^2 - 1}\right) = 2K_0(k_t b), \tag{S9}$$

where we introduced the transverse wave vector of the electron field $k_t \stackrel{\text{def}}{=} \omega/(\gamma_h v)$. The last transformation in Eq. (S9) remains valid on the boundary of the first quadrant (for $\text{Im}\,\beta_h = 0$), specifically, when $\beta_h > 0$ (but $\beta_h \neq 1$), because both $\gamma_h$ (or $k_t$) and the principal branch of $\sqrt{\beta_h^2 - 1}$ are continuous with respect to $\beta_h$ approaching the boundary from inside the quadrant.

Therefore, the validity of the whole Eq. (S9) follows from the analytic continuation of the integral to the boundary of the quadrant.

However, for additional rigor, we further separately consider the cases of $0 < \beta_h < 1$ and $\beta_h > 1$. In the former case (including the case of vacuum) we introduce $w \in \mathbb{R}$ such that

$$\tanh w = \beta_h \Rightarrow \sinh w = \beta_h \gamma_h, \cosh w = \gamma_h. \tag{S10}$$

Then collecting $\sinh(u - w)$ inside the exponent and replacing $u - w \to u$, we obtain

$$I_S(b) = \int_{-\infty}^{\infty} du \exp(-ik_t b \sinh u) = 2K_0(k_t b), \tag{S11}$$

where the latter equality follows from Eq. 10.32.6 of [32]. For the case $\beta_h > 1$ we introduce $w \in \mathbb{R}$ such that

$$\coth w = \beta_h \Rightarrow \sinh w = -i\gamma_h, \cosh w = -i\beta_h \gamma_h. \tag{S12}$$

Then collecting $\cosh(u - w)$ we obtain similarly to Eq. (S11)

$$I_S(b) = \int_{-\infty}^{\infty} du \exp(-k_t b \cosh u) = \pi i H_0^{(1)}(ik_t b) = 2K_0(k_t b), \tag{S13}$$

where the integral representation of $H_0^{(1)}$ is based on Eq. 10.9.9 of [32] (since in this case $ik_t > 0$). Combining the expression for $I_S(b)$ with Eq. (S7), we finally obtain:

$$I_1(\mathbf{r}) = \frac{q}{2\pi} \exp\left[i\frac{\omega}{v}(z - z_0)\right] K_0\left(\frac{\omega b}{\gamma_h v}\right). \tag{S14}$$

## S4 Ultra-relativistic scaling

Let us, first, discuss the behavior of the incident field in the case $\omega b \ll |\gamma_h| v$, when Eq. (25) simplifies into

$$\mathbf{E}_{\text{inc}}(\mathbf{r}) \approx \frac{q}{2\pi\varepsilon_h v} \exp\left[i\frac{\omega}{v}(z - z_0)\right] \frac{1}{b^2} \begin{pmatrix} x - x_0 \\ y - y_0 \\ 0 \end{pmatrix}. \tag{S15}$$

Here we used Eq. 10.31.1 of [32] and the fact that $-\ln b \ll 1/b$. For moderately fast electrons ($|\gamma_h| \sim 1, \beta_h \sim 1$) Eq. (S15) is valid whenever $b$ is much smaller than the light wavelength in the medium, i.e. in the classical quasi-static regime ($kb \ll 1$). If $|z - z_0|$ is comparable to $b$, then we can neglect the remaining exponent as well leading to the field of a wire with static current (independent of $\omega$). However, for $v \to c/m_h$ we have $|\gamma_h| \gg 1$; then Eq. (S15) remains valid for much larger $b$ with extra suppression of $E_{\text{inc},z}$. In other words, ultrarelativistic electron (in dense media) always leads to quasi-static dependence of the field versus transverse coordinates. Importantly, this regime shows that $\mathbf{E}_{\text{inc}}$ is everywhere continuous with respect to $\beta_h$ (or $m_h$) and have a finite (and simple) limit for $\beta_h \to 1$. However, its derivative with respect to $\beta_h$ has logarithmic singularity at this point.

The rigorous scaling, described in Section 2.7, breaks down when the new value of $v$ is larger than $c$. More specifically, the general scaling of Eqs. (43), (44) works perfectly well in the Cherenkov case, but then none of the two equivalent host media can be a vacuum one (due to $\beta_h > 1$). However, we can try to push it a bit further using the quasi-static limit of Eq. (S15). Assuming the exponent to weakly depend on $v$ (apart from a constant phase factor, which is irrelevant for computed probabilities), we obtain:

$$v \to c, \qquad \mathbf{E}_{\text{inc}} \to \beta_h \mathbf{E}_{\text{inc}}, \qquad P_{\text{EELS,CL}} \to \beta_h^2 P_{\text{EELS,CL}}. \tag{S16}$$

Thus, we may use any of the two scaling approaches above, then set $v = c$ (corresponding to $\mathcal{E} \gg \mathcal{E}_0$) in vacuum simulation, and additionally divide the obtained probabilities by $\beta_h^2$. Note, however, that this approximation does not correctly describe the first order of deviation of probabilities when $\beta_h$ increases from 1 due to logarithmic singularity of the derivative discussed in Section 2.2.

## S5 Code implementation

To perform EELS and CL simulations according to the developed general theory, we modified the open-source software ADDA.[21] First, we added an option to use the electric field of a relativistic electron [Eq. (25)] as the incident field in ADDA. A user has to specify the electron kinetic energy in keV and coordinates of the beam (trajectory) center $(x_0, y_0, z_0)$ in the laboratory reference frame. If needed, propagation direction can be changed from the default one (along the $z$-axis). Note, that moving the beam center along the electron trajectory (e.g., changing $z_0$ for the default propagation direction) affects only the constant phase factor for both incident and total fields, but none of the measurable quantities. ADDA now has an option to specify arbitrary host-medium refractive index $m_h \in \mathbb{C}$, although the standard light-scattering problem for a particle in an absorbing host medium is still a field of active research.[28,43,57]

A technical complication arises from the fact that currently ADDA is based on the Gaussian-CGS system of units in contrast to SI used in this manuscript. Moreover, the unit of length is assumed to be nm in contrast to any other electromagnetic excitation (when any unit of length can be used as long as it is the same for all input and output quantities).[44] Therefore, the incident field given by Eq. (25) is additionally transformed from V·s/m into statV·s/cm in ADDA, which then computes all cross sections including $C_{\text{enh}}$ [Eq. (38)] in nm$^2$, assuming $E_0 = 1$ statV·s/cm. It further computes $P_{\text{EELS}}$ and $P_{\text{CL}}$ using Eqs. (37) and (41), respectively, and identity $4\pi\varepsilon_0 E_0^2 = 0.1$ J·s$^2$/m$^3$.

Particle refractive index $m_\text{p}$ (or its distribution inside the particle) depends on the wavelength $\lambda$ corresponding to the desired energy loss $\hbar\omega$. A single run of ADDA performs a simulation for a single set of parameters ($\lambda$, beam position, etc.). To simulate a loss spectrum or to scan a particle's cross-section with the beam, it is necessary to run ADDA multiple times varying the desired parameters. For instance, to simulate the EELS and CL spectra one needs to vary $\lambda$ and the corresponding $m_\text{p}$. We developed a Python wrapper, named `ADDAwrapper`, to automate this process.

The wrapper is designed to work as a Python library, so one only has to fill the example preset file with the simulation parameters, and call high-level functions from this file to perform the corresponding set of simulations (spectrum, loss probability scan over the cross-section of the particle, etc.), collect the data from these simulations, and plot it in various formats. Multithreading is supported to run several single-thread ADDA runs simultaneously on different processor cores. Such use of ADDA in sequential mode is both easier and more efficient than the use of the parallel ADDA mode (based on Message Passing Interface), when run on a shared-memory machine and memory is not a limiting factor. In this case, the script leads to a speedup equal to the number of physical processor cores without any efforts from a user. The wrapper is distributed with the ADDA code in the folder `/misc/ADDAwrapper`.

ADDA supports different options for computational procedures: polarizability prescription, iterative solver, interaction term, etc.[44] An optimal set of these options depends on the specific light scattering problems. From our experience, the best speed with satisfactory precision for EELS and CL simulations of nanoparticles are achieved with the following set of options. Iterative solver QMR2 (quasi-minimal residual method for complex symmetric matrices based on a two-term recurrence) works slightly faster than the default QMR (based on a three-term recurrence) for the same precision. Polarizability prescription and interaction term based on integration of Green's tensor ("`-pol igt_so`" and "`-int igt_so`", respectively) give more precise results than all other options, and are faster for particles with higher surface/volume ratio. Setting the stopping threshold (relative residual) of the iterative solver $\varepsilon_\text{it}$ to $10^{-2}$ (instead of the default value $10^{-5}$) by the command line option "`-eps 2`" accelerates the simulation several times keeping the satisfactory accuracy (errors due to the iterative solver are about 1%). Such accuracy level is generally acceptable for DDA simulations of metallic nanoparticles, since improving it requires very fine discretization in the first place.[52] Apart from multithreading, using the above parameters allows the wrapper to perform EELS and CL simulations with ADDA up to an order of magnitude faster than with default code settings. The updated code is available at https://github.com/alkichigin/adda and will be implemented into official ADDA release soon.

# S6 Additional simulations: cathodoluminescence and transition to the Cherenkov case

CL probabilities simulated in ADDA for different refractive indices of the host medium.

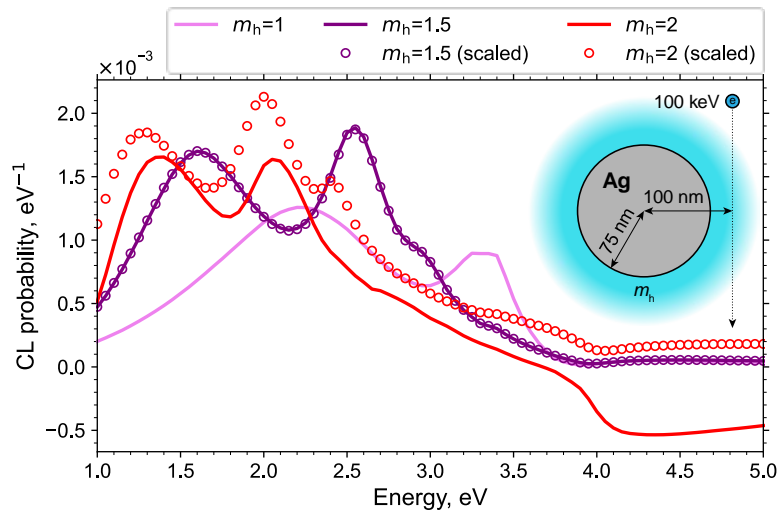

Fig. S1. CL spectra for a sphere in an infinite non-absorbing host medium simulated with ADDA ($n_x = 128$), either directly or through scaling of simulations in vacuum (the same as Fig. 5, but for CL). The problem parameters are shown in the inset. In the host medium with $m_h = 2$ the electron is faster than light, causing the Cherenkov radiation.

Next, we study how $P_{EELS}$ and $P_{ext}$ deviate from each other when $m_h$ increases over the Cherenkov threshold (Fig. S2). First, one can see the inflection at this threshold ($m_h = c/v$), corresponding to the logarithmic singularity of the derivative discussed in Section 2.2. Second, we have also used the approximate scaling of Eq. (S16). As expected, it gives the wrong value of derivative near the inflection point. But, surprisingly, it describes the general trend of $P_{EELS}$ for larger values of $m_h$, which is very different from that of $P_{ext}$. This partly explains the satisfactory performance of this scaling in Fig. 5 and Fig. S1. The specific reasons for that are still not clear.

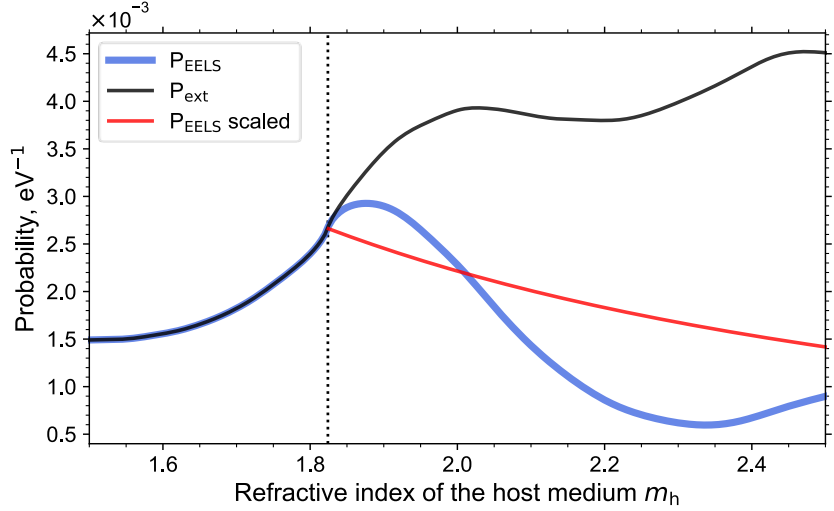

Fig. S2. EELS and extinction probabilities of 2.0 eV energy loss simulated with the DDA ($n_x = 32$) for different host-medium refractive indices. The parameters of the problem are the same as in Fig. S1.

The following figures show the comparison of CL results computed with ADDA, the Lorenz–Mie theory, and MNPBEM in different host media. All three methods agree for the case of vacuum (Fig. S3), but for $m_h = 1.5$ only ADDA and Lorenz–Mie theory agree (as in Fig. S1), while MNPBEM results are markedly different. We have, however, took the liberty to multiply the latter by $m_h$; this restores the agreement. We have not investigated this issue in details, but the error may be due to the fact that the BEM calculates $P_{CL}$ through a far-field integral (similar to $C_{sca}$ in ADDA), which implicitly depends on the definition of $I_0$ containing $m_h$ [Eq. (19)]. Practically, this difference is not important since all CL measurements are performed in arbitrary (relative) units. Still, the definition of $P_{CL}$ by Eq. (42) is the correct one, since, e.g., it guarantees that $P_{CL} = P_{EELS}$ for the case when both a particle and a host medium are non-absorbing (and $\beta_h < 1$).

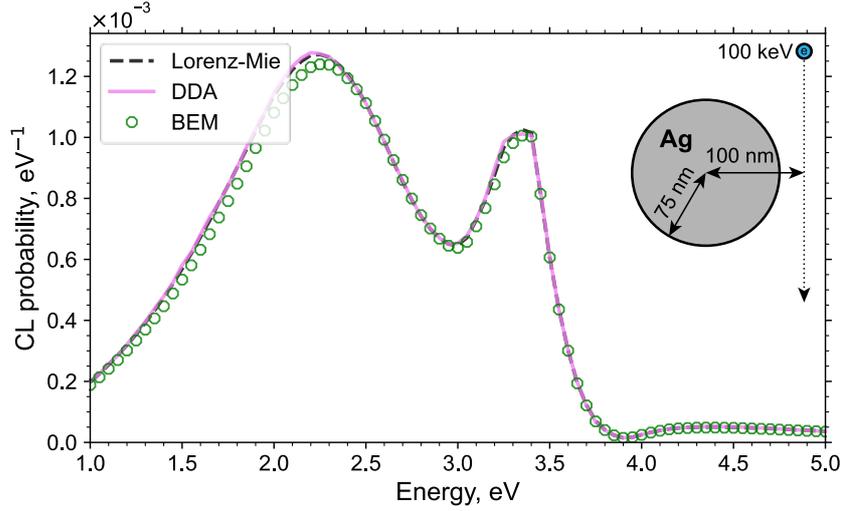

Fig. S3. CL spectra simulated with the Lorenz-Mie theory, ADDA ($n_x = 128$ with the extrapolation), and the BEM (the same as Fig. 6, but for CL). The parameters of the problem are shown in the inset.

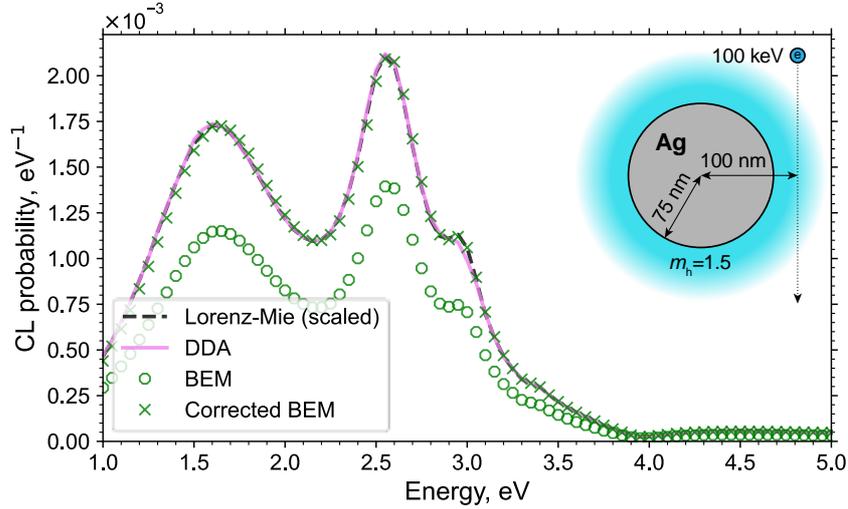

Fig. S4. CL spectra simulated with the Lorenz-Mie theory (using scaling), ADDA ($n_x = 128$ with the extrapolation), and the BEM (the same as Fig. 7, but for CL). The corrected BEM results are obtained multiplying by $m_h$. The parameters of the problem are shown in the inset.

Similar scaling issue can be seen for the Cherenkov case ($m_h = 2$) in Fig. S5, but here we additionally depict $P_{\mathrm{CL}}^{\mathrm{tot}}$, defined by Eq. (41). Apart from the overall magnitude, it has certain qualitative difference from $P_{\mathrm{CL}}$. For instance, the peak around 2.5 eV has almost disappeared.

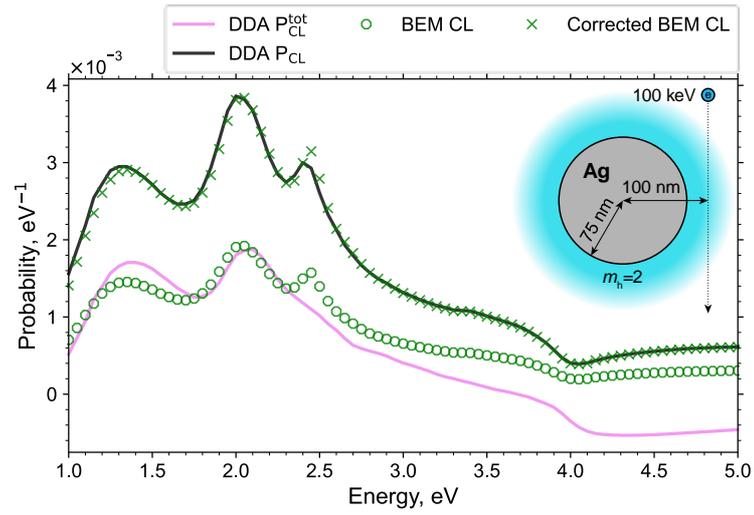

Fig. S5. CL probability simulated with the DDA ($n_x = 128$ with the extrapolation) and the BEM (the same as Fig. 8, but for CL). For the DDA both $P_{\mathrm{CL}}$ and $P_{\mathrm{CL}}^{\mathrm{tot}}$ are shown, the corrected BEM results are obtained multiplying by $m_h$. The parameters of the problem are shown in the inset.